%% file: main.tex
\documentclass[sigconf]{acmart}
\usepackage{graphicx}
\usepackage[tikz]{bclogo}
\usepackage{syntax}
\usepackage{amsmath,amsfonts}
\usepackage{algorithmic}
\usepackage{textcomp}
\usepackage[table]{xcolor}
\usepackage{pgfplots}
\usepackage{pgfplotstable}
\usepackage{threeparttable}
\usepackage{subcaption}
\usepackage{xcolor}
\usepackage{hyperref} 
\usepackage{url}
\usepackage{booktabs}
\usepackage{csquotes}
\usepackage{adjustbox}
\usepackage{standalone}
\usepackage{tikz}
\usepackage{bm}
\usepackage{multirow}
\usepackage{subcaption}
\usepackage{tcolorbox}

\usepgfplotslibrary{fillbetween}

\mathchardef\mhyphen="2D

\DeclareMathAlphabet\mathbfcal{OMS}{cmsy}{b}{n}

\DeclareMathOperator*{\argmax}{\textbf{arg\,max}}

\newcommand{\approach}{\texttt{LQPR}}

{\noindent\begin{minipage}[c]{\linewidth}%
\begin{bclogo}[couleur=white,%
                arrondi=0.1,%
                ombre=true]{} {#1}}%
{\end{bclogo}\end{minipage}\vspace{2mm}}

{\noindent\begin{minipage}[c]{\linewidth}%
\begin{bclogo}[couleur=gray!20,%
                arrondi=0.1,%
                logo=\bclampe,%
                ombre=true]{\normalsize #1} {#2}}%
{\end{bclogo}\end{minipage}\vspace{2mm}}

\def\BibTeX{{\rm B\kern-.05em{\sc i\kern-.025em b}\kern-.08em
    T\kern-.1667em\lower.7ex\hbox{E}\kern-.125emX}}

\newcommand{\keystate}[1]{
\begin{tcolorbox}[colback=gray!20,leftrule=1mm,toprule=0mm,bottomrule=0mm,left=1pt,right=2pt,top=2pt,bottom=2pt]
\em #1
\end{tcolorbox}
}

\newcommand{\sta}[1]{
\begin{tcolorbox}[colback=gray!20,leftrule=0mm,rightrule=0mm,toprule=0mm,bottomrule=0mm,left=0pt,right=0pt,top=1pt,bottom=1pt]
\em #1
\end{tcolorbox}
}

\newcommand{\revisioncolor}{black} 
\newcommand{\revision}[1]{{\color{\revisioncolor}#1}}

\newcommand{\finalrevisioncolor}{black} 
\newcommand{\finalrevision}[1]{{\color{\finalrevisioncolor}#1}}

\copyrightyear{2026}
\acmYear{2026}
\setcopyright{cc}
\setcctype{by}
\acmConference[ICSE '26]{2026 IEEE/ACM 48th International Conference on Software Engineering}{April 12--18, 2026}{Rio de Janeiro, Brazil}
\acmBooktitle{2026 IEEE/ACM 48th International Conference on Software Engineering (ICSE '26), April 12--18, 2026, Rio de Janeiro, Brazil}
\acmPrice{}
\acmDOI{10.1145/3744916.3773107}
\acmISBN{979-8-4007-2025-3/26/04}

\begin{document}

\begin{abstract}

Elicited performance requirements need to be quantified for compliance in different engineering tasks, e.g., configuration tuning and performance testing. Much existing work has relied on manual quantification, which is expensive and error-prone due to the imprecision. In this paper, we present \approach, a highly efficient automatic approach for performance requirements quantification. \approach~relies on a new theoretical framework that converts quantification as a classification problem. Despite the prevalent applications of Large Language Models (LLMs) for requirement analytics, \approach~takes a different perspective to address the classification: we observed that performance requirements can exhibit strong patterns and are often short/concise, therefore we design a lightweight linguistically induced matching mechanism. We compare \approach~against nine state-of-the-art learning-based approaches over diverse datasets, demonstrating that it is ranked as the sole best for 75\% or more cases with two orders less cost. Our work proves that, at least for performance requirement quantification, specialized methods can be more suitable than the general LLM-driven approaches.

\end{abstract}

\title{Light over Heavy: Automated Performance Requirements Quantification with Linguistic Inducement}

\author{Shihai Wang}
\email{wsh2130076635@gmail.com}
\authornote{Shihai Wang is also supervised in the IDEAS Lab.}
\affiliation{%
  \institution{School of Computer Science and Engineering\\University of Electronic Science and Technology of China}
  \city{Chengdu}
  \country{China}
}

\author{Tao Chen}
\email{t.chen@bham.ac.uk}
\authornote{Tao Chen is the corresponding author.}
\affiliation{%
  \institution{IDEAS Lab, School of Computer Science\\University of Birmingham}
  \city{Birmingham}
  \country{UK}
}



\maketitle

\input{introduction}

\input{background}
\input{empirical}

\input{theory}

\input{method}

\input{study}

\input{results}

\input{discussion}

\input{threat}

\input{related}

\input{conclusion}

\section*{Acknowledgment}
This work was supported by a NSFC Grant (62372084) and a UKRI Grant (10054084).

\balance

\bibliographystyle{ACM-Reference-Format}
\bibliography{Mybib}
\citestyle{acmauthoryear} 
\end{document}

%% file: introduction.tex
\section{Introduction}
\label{sec:intro}

Failure to comply with the requirements on the behavioral quality of software systems, such as performance requirements, is often a major reason of unsuccessful software projects~\cite{DBLP:conf/re/EckhardtVFM16,DBLP:journals/tse/SayaghKAP20,DBLP:journals/tosem/ChenL23a,abdeen2023approach,DBLP:conf/icse/ChenChen26,DBLP:conf/icse/XiongChen25,DBLP:conf/icse/XiangChen26}. The root cause is that those requirements, even after a formal elicitation, remain difficult to interpret and quantify. For example, consider two elicited performance requirements:
\sta{
\centering
\textit{``{The search shall take no longer than 15 seconds.}''}
} 
\sta{
\centering
\textit{``{The search shall return in 15 seconds.}''}
} 
At the first glance, those two requirements appear to be almost identical. However, according to existing work~\cite{DBLP:conf/re/EckhardtVFM16,DBLP:journals/tse/SayaghKAP20} and after the consultation with our industry partners, their interpretations and the preferences implied can be rather different: the former implies that anything longer than 15 seconds is not acceptable since ``\texttt{no longer than}'' is such a strong term thereof while, imprecisely, there could be some preferences for latency smaller than 15 seconds. The latter, in contrast, implies that anything better than 15 seconds is equally preferred, and there may be certain tolerances when the performance fails below the expectation. 

Understanding the above interpretation directly determines how to quantify the performance requirements. This quantification is important, e.g., it has been prone that depending on how the performance requirements are quantified in guiding the configuration tuning, the achieved performance can vary considerably~\cite{DBLP:journals/tosem/ChenL23a}. In performance testing, the quantification is also fundamental for the oracle that determines when a ``performance bug'' is detected~\cite{DBLP:conf/icse/MaChen25}.


Yet, manual quantification is expensive: there could be hundreds of performance requirements for a software project~\cite{DBLP:conf/icse/EckhardtVF16}. More importantly, quantifying the requirements manually requires software engineers to comprehend the meaning, reasoning about them based on domain knowledge before making inference. Each of those steps, if not done correctly, can result in subjectively biased and misleading conclusions.


While domain specific languages exist for formalizing requirements~\cite{DBLP:journals/re/WhittleSBCB10,DBLP:conf/re/BaresiPS10}, they lack holistic, mathematical and generalizable ways to quantify performance requirements. Manual analysis is often required therein to parse, reason, and construct correct formal specifications, which is labor-intensive and challenging. Indeed, since requirements are \revision{written in} natural languages, leveraging machine learning, especially Large Language Models (LLMs), have been dominating for requirement analysis~\cite{DBLP:journals/entropy/CanedoM20,DBLP:journals/tse/ZhaoXW24,DBLP:conf/re/DalpiazDAC19,DBLP:conf/naacl/DevlinCLT19,DBLP:journals/corr/abs-1907-11692,mayer2023prompt,DBLP:conf/emnlp/SunL0WGZ023}. Yet, in this work we demonstrate a \textbf{\textit{``light over heavy''}} phenomenon: those \textbf{\textit{``heavy''}} and general learning-based approaches can be much inferior to a \textbf{\textit{``light''}} yet tailored approach according to domain understandings of the problem characteristics.

In this paper, we propose \approach, a holistic framework that automatically quantifies performance requirements. For the first time, we formulate the quantification of performance requirements as a classification problem within a formal theoretical framework. Unlike the learning-based approaches, what make \approach~unique is that it is lightweight, induced by linguistic knowledge that is specifically tailored to fit our empirical observations and the newly formulated classification. Our key contributions are:

\begin{itemize}
    \item An empirical study that discloses common characteristics for 259 real-world performance requirements (Section~\ref{sec:emprical}).
    \item Drawing on the empirical observations, we formulate a theoretical framework that converts the requirement quantification as a classification problem (Section~\ref{sec:theory}).
    \item We design a lightweight, linguistically induced solution to automatically classify/quantify performance requirements with dually syntactic and semantic scoring (Section~\ref{sec:method}).
    \item We compare \approach~against \revision{10} state-of-the-art approaches, including LLMs, for requirements engineering under diverse datasets and metrics (Section~\ref{sec:study}).
\end{itemize}

The results from Section~\ref{sec:exp} are encouraging: \approach~outperforms other state-of-the-art over all datasets, in which 11 out of 15 cases it is statistically ranked as the sole best. \approach~does so under little resources/overhead that is around two orders less than using LLMs.


A key takeaway of this work is: while pre-trained models like LLMs are dominating for software/requirement engineering, we show that at least for cases like performance requirements quantification in which strong domain-specific problem formulation is required, there exist simpler while much more efficient solutions, hence we urge the community to take a step back when dealing with a similar software engineering case in the current LLM era. All data is published at: \textcolor{blue}{\texttt{\url{https://github.com/ideas-labo/LQPR}}}.

For other major organization, Sections~\ref{sec:pre},~\ref{sec:discussion}, and~\ref{sec:con} present the scope/problem, discussions, and conclusion, respectively.

%% file: background.tex
\section{Preliminaries}
\label{sec:pre}

\subsection{Context and Scope}


Performance requirements serve as the stakeholders' aspiration to the behavioral quality of software systems. In this work, we focus on the requirements that have gone through \revision{standardized} requirement elicitation procedure rather than those from public platforms, such as \texttt{StackOverflow}. Commonly, the requirements are elicited into different statements in the documentation, which can still often of complex implied preferences and high imprecision~\cite{DBLP:conf/re/EckhardtVFM16}.

\subsection{Problem Formulation}

The inherently implied preferences and imprecision from the performance requirements cause great challenges to many software engineering tasks, which often require precise definition and quantification on the performance needs. Manual interpretation of performance requirements and their implication is not only labor-intensive, but also error-prone, causing devastating consequences.

For example, Chen and Li~\cite{DBLP:journals/tosem/ChenL23a} have demonstrated that the performance requirements and their satisfactions can significantly influence configuration tuning if used therein, leading to better compliance than tuning without. However, unrealistic requirements could be harmful. For performance testing, incorrectly quantifying them as the satisfaction of performance can lead to flawed oracle, causing severe resource waste~\cite{DBLP:conf/kbse/HeJLXYYWL20,DBLP:conf/icse/MaChen25}. Yet, in another example of self-adapting the systems, misrepresenting performance needs can trigger excessive adaptations or limit the adaptability~\cite{DBLP:journals/tsc/ChenB17,DBLP:conf/icse/YeChen25}. 



Therefore, the problem is to automatically build the following
\begin{equation}
\mathbfcal{R} \rightarrow g \text{; }s = g(v)  
\end{equation}
whereby $v$ is a concerned performance value and $s$ is the corresponding satisfaction interpreted from the statement of performance requirement $\mathbfcal{R}$. The core is how to build the function $g$ that automatically interprets a given $\mathbfcal{R}$, which is the focus of this work.

%% file: empirical.tex
\section{Understanding Performance Requirements}
\label{sec:emprical}

\finalrevision{To understand the state-of-the-practice for documenting software performance requirements, we conduct an empirical study on the requirements from the \textsc{Promise}~\cite{promise}---a widely-used dataset~\cite{DBLP:journals/infsof/AlhoshanFZ23,DBLP:conf/re/DalpiazDAC19} that contains software engineering project data primarily from academic settings\footnote{While other datasets exist, e.g., see Section~\ref{subsec:dataset}, here we focus on the \textsc{Promise} since it has one of the largest set of requirement samples; further, we will evaluate the understandings generalized from \textsc{Promise} in Section~\ref{sec:exp}.}.} \revision{The process of deriving the performance requirements from the dataset is as follows:}
\revision{
\begin{enumerate}



\item Screen the requirements using keywords related to performance, e.g., \texttt{``fast''}, \texttt{``low''}, and \texttt{``timeout''}. A requirement is a candidate if it matches any of the keywords.

        \item Any requirements that contain numbers, or words that are quantifiable, e.g., \texttt{``all''}, are also candidates.

        \item Verify the candidates: those that contain performance keywords and are quantifiable will be selected. Otherwise, we manually confirm if the requirement is relevant since it is possible for a performance requirement without expectation, e.g., \texttt{``the system shall be fast''}, or vice versa.
        
\end{enumerate}}

This has led to 212 unique performance requirements. 



\begin{figure}[t!]
\centering
\begin{minipage}{0.45\columnwidth}
\centering
\resizebox{1.1\textwidth}{!}{

\input{tables/new-ep1}
}
\vspace{0.6cm}
\subcaption{Performance expectations.}
\label{tab:rq2} 
\end{minipage}
~\hspace{0.3cm}
\begin{minipage}{0.45\columnwidth}
\centering
\resizebox{1.1\textwidth}{!}{
\input{tables/new-ep2}
}
\vspace{0.22cm}
\subcaption{Imprecision types.}
\label{tab:rq2} 
\end{minipage}
\vspace{-0.3cm}
  \caption{Statistics of the empirical study. In (b), a requirement might belong to more than one imprecision type.}
  \label{fig:ep}
\end{figure}

\subsection{Prevalence of Performance Expectations}

A noticeable phenomenon across all the performance requirements is that it often contains a number that represents some information of preferences, which we call \textbf{expectation point}. Overall, Figure~~\ref{fig:ep}a shows that up to 61.3\% of the performance requirements have one expectation point, such as:
\sta{
\centering
\textit{``{The system shall support at least 1,000 users.}''}
}
Those minority of requirements with more than one expectation points can be easily decomposed/split, e.g., the requirement:

\sta{
\centering
\textit{``{The system shall react in 5 seconds and ideally less than 2 seconds.}''}
}
\noindent can be trivially broken down as:
\sta{
\centering
\textit{``{The system shall react in 5 seconds.}''}
}
\sta{
\centering
\textit{``{The system shall react ideally less than 2 seconds.}''}
}
\noindent The above can then be processed and quantified separately\footnote{Note that the two expectation points should describe the same condition, subject, or action; or otherwise only one (often the latter) is the true expectation.}. We also see that 16.5\% requirements do not have expectation, which naturally prefer a best possible outcome of the performance. e.g., 
\sta{
\centering
\textit{``{The system shall be fast.}''}
}

Even though an explicit expectation point is specified, the requirement itself can still be largely imprecise and hard to quantify. 


\keystate{
\textit{\textbf{Finding 1}: Up to 83.5\% performance requirements have expectation point(s), of which 61.3\% contains exactly one.}
}

\subsection{Imprecision in Performance Requirements}

Next, we split those performance requirements with two expectation points into one each, which leads to 259 requirements, and classify their types of imprecision as the categories below:

\begin{itemize}
\item \textbf{Type-I:} It is not clear about the level of tolerance for results worse than the expectation. 
\item \textbf{Type-II:} It is not clear about the extent of preference for results better than the expectation. 
\item \textbf{Type-III:} It is not clear to what extent is sufficient.
\item \textbf{N/A:} The requirement has no imprecision involved.
\end{itemize}

For example, ``\texttt{\texttt{The system should be fast}}'' would fail into \textbf{Type-III} while ``\texttt{\texttt{The system should support at least 1,000 users}}'' would fit \textbf{Type-II}, as ``\texttt{at least}'' is such a strong phrase that indicates something is necessary or required, i.e., no throughput less than $1,000$ is acceptable. However, it remains unclear about to what extent throughput better than $1,000$ are preferred. The categorization results are presented in Figure~~\ref{fig:ep}b.


\keystate{
\textit{\textbf{Finding 2}: \revision{All performance requirements contain imprecision, and the majority of them are missing one end of the scale (\textbf{Type-I} or \textbf{Type-II}).}}
}

\subsection{Insights}
\label{sec:insights}

\revision{Drawing on the findings, we derive the following insights:}
\revision{
\begin{itemize}
    \item Performance requirements typically exhibit only one threshold point. This implies that when transformed into quantitative functions, the possible shapes of these functions are relatively limited, suggesting the high potential of constructing more general patterns (\textbf{\textit{Finding 1-2}}).
    
    \item The inherent ambiguity in performance requirements presents a significant challenge (\textbf{\textit{Finding 2}}).

\end{itemize}
}

%% file: tables/new-ep1.tex
\begin{tabular}{|lll|}
\toprule
\textbf{$\#$ Expectations Points} & {\textbf{Count}} &  {\textbf{\%}}\\ 
\midrule

No expectation&35&16.5\%\\
One expectation&130&61.3\%\\
Two expectations&47&22.1\%\\
\bottomrule
\end{tabular}

%% file: tables/new-ep2.tex
\begin{tabular}{|lll|}
\toprule
\textbf{Imprecision Type} & {\textbf{Count}} &  {\textbf{\%}}\\ 
\midrule

Type-I&108&41.7\%\\
Type-II&126&48.6\%\\
Type-III&35&13.5\%\\
N/A&0&0\%\\

\bottomrule
\end{tabular}

%% file: theory.tex
\section{Theoretical Framework in \approach}
\label{sec:theory}

\subsection{Fragments with Implied Preferences}

Since the performance requirements can be patternized based on the number of expectation points (\textbf{\textit{Finding 1}}) and there are different imprecision types (\textbf{\textit{Finding 2}}), in \approach, we formally specify them with fragments as these points essentially represent distinct preferences based on the intervals of metric values. Given an interval $[v_i,v_{i+1}]$ over the value $v$ of a performance metric $y$, the fragment and its implied preference of the requirement, denoted as $\psi$, can be represented as Backus-Naur notations~\cite{DBLP:journals/cacm/Knuth64a} in \approach~below:
\sta{
\vspace{0.3cm}
\begin{grammar}
<$\psi$> ::=  $\mathcal{G}$ | $\mathcal{S}$ | $\mathcal{E}$
\end{grammar}
\begin{grammar}
<$\mathcal{G}$> ::= $\forall v \in [v_i,v_{i+1}]$, a greater $v$ is preferred at $[s_i,s_{i+1}]$
\end{grammar}
\begin{grammar}
<$\mathcal{S}$> ::= $\forall v \in [v_i,v_{i+1}]$, a smaller $v$ is preferred at $[s_i,s_{i+1}]$
\end{grammar}
\begin{grammar}
<$\mathcal{E}$> ::= $\forall v \in [v_i,v_{i+1}]$ is equally preferred at $s_i$
\end{grammar}
}
\noindent where $s_i$ denotes the satisfaction score for that interval ($s_i \in [0,1]$), which is adapted depending on the preference of the adjacent intervals in a performance requirement. The first two are \textbf{distinguishable} fragments while the last is an \textbf{indistinguishable} fragment. Clearly, a score of 1 and 0 represent fully satisfied and fully non-satisfied requirement, respectively. Note that while some bounds for a performance metric are clear, e.g., the lower bound for latency and throughput can only be $0$, in other cases, the bounds may be unknown, e.g., the maximum latency. Therefore, mathematically, we allow $v_i = -\infty$ and/or $v_{i+1} = \infty$ as needed.

\begin{figure}[!t]
 \centering
  \begin{subfigure}[t]{0.35\columnwidth}
    \centering
\includegraphics[width=\textwidth]{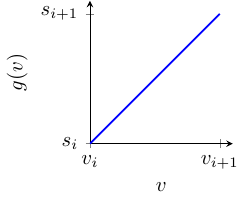}
  \subcaption{$\mathcal{G}$}
  \end{subfigure}
 ~\hspace{-0.3cm}
  \begin{subfigure}[t]{0.35\columnwidth}
    \centering
\includegraphics[width=\textwidth]{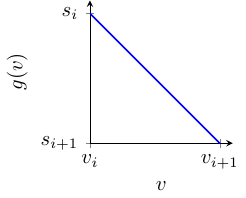}
    \subcaption{$\mathcal{S}$}
  \end{subfigure}
 ~\hspace{-0.3cm}
  \begin{subfigure}[t]{0.35\columnwidth}
    \centering
\includegraphics[width=\textwidth]{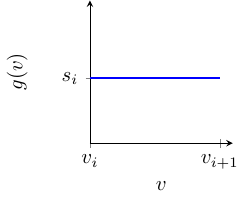}
   \subcaption{$\mathcal{E}$}
  \end{subfigure}
  
    \caption{Fuzzy functions $g(v)$ that quantifies the satisfaction score of all three types of fragment. (a): $g(v)=({{s_{i}-s_{i+1}}\over{{v_i-v_{i+1}}}})v + {{s_{i+1}v_i- s_{i}v_{i+1}}\over{{v_i-v_{i+1}}}}$; (b): $g(v)=({{s_{i}-s_{i+1}}\over{{v_i-v_{i+1}}}})v + {{s_{i+1}v_i- s_{i}v_{i+1}}\over{{v_i-v_{i+1}}}}$; (c): $g(v)=s_i$.}
   \label{fig:fragments}
   \vspace{-0.3cm}
 \end{figure}

The satisfaction score of the above fragments can be quantified by a function $g(\cdot)$ using fuzzy logic~
\cite{zadeh1988fuzzy}. As shown in Figures~\ref{fig:fragments}a and~\ref{fig:fragments}b, since we do not know to what extent a greater (or smaller) $v$ is sufficient, the corresponding membership function can be linearly specified as a slop that monotonically increases (or decreases) the satisfaction score gradually from $v_i$ to $v_{i+1}$. Both fragments reach the best and worst satisfaction at the bounds of the interval\footnote{For maximizing metric, $s_i \leq s_{i+1}$; otherwise, $s_i \geq s_{i+1}$.}. The fragment ``$\forall v \in \theta$ is equally preferred at $s_i$" is essentially a special case of the fuzzy logic, such that all values of the performance metric between $v_i$ and $v_{i+1}$ are equally satisfied at $s_i$ (Figure~\ref{fig:fragments}c).

\subsection{Quantifiable Propositions of Requirements}

In theory, the fragments can be arbitrarily combined, in any number or order, to form a complex yet quantifiable proposition for the performance requirement. Any two fragments are joint by an expectation point\footnote{Note that, theoretically we can have two fragments joint by an expectation point, where both of them have monotonically decreasing, increasing or consistent satisfactions, but it has no practical meaning, since practically it can be reduced to as having a single decreasing, increasing or consistent satisfactions, and hence it is ruled out from consideration. Yet, this can be a symbolic class label as we will discuss.}. Formally, with the Backus-Naur notations in \approach, a proposition $p$ of $n$ fragments ($n-1$ expectation points) is:
\sta{
\vspace{0.3cm}
\begin{grammar}
\centering
<p> ::= $\psi$ | $\psi$ \& $p$
\end{grammar}
}
\noindent in which there are $n$ intervals, i.e., $\{[v_1,v_2], [v_2,v_3]$ $,...,$ $[v_n,v_{n+1}]\}$, and a vector of satisfaction scores, e.g., $\boldsymbol{\overline{s}}=\{[s_1,s_2],s_2,...,$ $[s_{n-1},s_{n}]\}$. 

\subsubsection{Setting Satisfaction Scores}
\label{sec:set-score}

When the first fragment is indistinguishable ($\mathcal{E}$), then $s_1=1$ and $s_1=0$ for minimizing and maximizing performance metric, respectively. If the first fragment is distinguishable, i.e., $\mathcal{G}$ and $\mathcal{S}$, then on all cases, for the former there is $s_1=0$ and for the latter we have $s_1=1$. For a proposition, there might be multiple series of fragments with different change traces on satisfactions. We set the satisfaction based on two cases:

\begin{itemize}
    \item For a series of $l$ fragments with \textbf{monotonic change} on the satisfactions, such that the distinguishable and indistinguishable fragments interleaving each other, the $i$th satisfaction score $s_i$ ($i \geq 2$) in the series can be:
    \begin{align}
     \begin{split}
      s_i =
      \begin{cases}
      s_{begin} - {{i - 1} \over d} \text{ if monotonically decreasing series} \\
      s_{begin} + {{i - 1} \over d} \text{ otherwise}\\
      \end{cases}
      \end{split}
      \label{eq:sat1}
\end{align}
$s_{begin}$ is the satisfaction score at the beginning of the series.

\item For a series with only consecutive \textbf{indistinguishable} fragments, we have $s_i = 1 - s_{i-1}$ ($i \geq 2$).
\end{itemize}




\subsubsection{Setting Intervals}
\label{sec:set-interval}

Generally, the intervals of two adjacent fragments $\psi_a$ and $\psi_b$ can be naturally combined without any changes. However, if they have the same interval $[v_{i},v_{i+1}]$ such that $\psi_a \neq \psi_b$ or their quantification differs, i.e., a conflict, we split the intervals as $[v_{i},{{v_i + v_{i+1}} \over 2}]$ and $[{{v_i + v_{i+1}} \over 2},v_{i+1}]$, where the former would have the quantification of $\psi_a$ while the later would use that of $\psi_b$. Otherwise, there is no need to change the intervals.







\begin{figure}[!t]
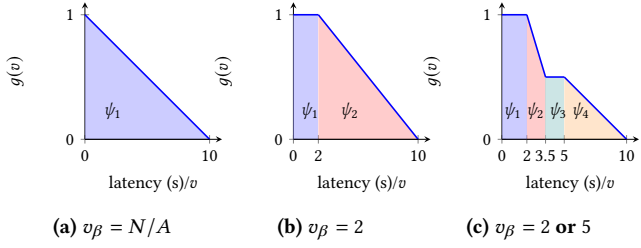

 \centering
  \begin{subfigure}[t]{0.35\columnwidth}
    \centering
\includestandalone[width=\textwidth]{figures/example1}
  \subcaption{$v_{\beta}=N/A$}
  \end{subfigure}
 ~\hspace{-0.3cm}
  \begin{subfigure}[t]{0.35\columnwidth}
    \centering
\includestandalone[width=\textwidth]{figures/example2}
    \subcaption{$v_{\beta}=2$}
  \end{subfigure}
 ~\hspace{-0.3cm}
    \begin{subfigure}[t]{0.35\columnwidth}
    \centering
\includestandalone[width=\textwidth]{figures/example3}
    \subcaption{$v_{\beta}=2\text{ or }5$}
  \end{subfigure}
  
    \caption{Quantification of exampled latency requirements with diverse $v_{\beta}$. Distinct fragments are colored differently.}
   \label{fig:perf}
  \vspace{-0.3cm}
 \end{figure}

\subsubsection{Examples}

Highly imprecise requirements like \texttt{``the system should be fast''} can be represented as $p = \psi_1$ (see Figures~\ref{fig:perf}a) with $v_1=0$ and $v_{2}=10$, suppose that we know the latency cannot be higher than $10$ seconds (e.g., due to the timeout setting). The satisfaction score $[s_1,s_2]$ would be $[1,0]$. This means that ``$\psi_1 = \mathcal{S} ::= \forall v \in [0,10]$, a smaller $v$ is preferred at $[1,0]$''.

Figure~\ref{fig:perf}b quantifies an example of one expectation point: ``\texttt{The system should response in 2 seconds}''. Here, there are two intervals $[v_1,v_2]=[0,2]$ and $[v_2,v_3]=[2,10]$ from both fragments. In \approach, this will be $p = \psi_1 \& \psi_2$. The implied preference therein is that anything better than 2 seconds is equally preferred while values greater than 2 seconds can be tolerated (as Section~\ref{sec:set-score}, the satisfaction score $s_1$ and $[s_1,s_2]$ are $1$ and $[1,0]$, respectively), then we have ``$\psi_1 = \mathcal{E} ::=  \forall v \in [0,2]$ is equally preferred at $1$'' and ``$\psi_2 = \mathcal{S} ::=  \forall v \in [2,10]$, a smaller $v$ is preferred at $[1,0]$''.

An example of two expectation points is given in Figure~\ref{fig:perf}c, where the requirement ``\texttt{The system should response in 5 seconds and ideally less than 2 seconds.}'' can be split into ``\texttt{The system should response in 5 seconds}'' and ``\texttt{The system should response in ideally less than 2 seconds.}'', each has implied preferences similar to that in Figure~\ref{fig:perf}b. Hence, according to Section~\ref{sec:set-score}, we now have $p = \psi_1 \& \psi_2\& \psi_3\& \psi_4$ where $s_1$, $[s_1,s_2]$, $s_2$, $[s_2,s_3]$ are $1$, $[1,0.5]$, $0.5$, $[0.5,0]$, respectively: ``$\psi_1 = \mathcal{E} ::=  \forall v \in [0,2]$ is equally preferred at $1$"; ``$\psi_2 = \mathcal{S} ::=  \forall v \in [2,5]$, a smaller $v$ is preferred at $[1,0.5]$"; ``$\psi_3 = \mathcal{E} ::=  \forall v \in [2,5]$ is equally preferred at $0.5$"; ``$\psi_4 = \mathcal{S} ::=  \forall v \in [5,10]$, a smaller $v$ is preferred at $[0.5,0]$". Yet, there is a conflict on the intervals for $\psi_2$ and $\psi_3$, hence as Section~\ref{sec:set-interval}, we set them as $[2,3.5]$ and $[3.5,5]$, respectively.

\subsection{Quantification as A Classification Problem}

It is easy to see that, propositions with multiple expectation points can be broken down into multiple ones with a single expectation point, while the ones without expectation are only special cases of those with a single expectation, i.e., with two identical fragments. Given this, and the fact that requirements with one expectation point form majority of the requirements, the key is (1) to classify the requirement with one (or none) expectation point in terms of the preference of each enclosed fragments using $\mathcal{G}, \mathcal{S}, \mathcal{E}$; and (2) to extract the expectation point $v_{\beta}$  (which might be N/A).

In \approach, we use a tuple $\langle\psi_l,\psi_r
\rangle$ to represent the classification label, where $\psi_l$ is for the fragment on the left while $\psi_r$ is the fragment on the right. In particular, $\langle\psi_l,\psi_r\rangle$ is represented by any combination of the aforementioned interpretation of $\mathcal{G}, \mathcal{S}, \mathcal{E}$. A requirement that has no expectation point is simply a special case of $\psi_l=\psi_r$. Therefore, the total classes number to be classified is $3^2=9$. Once a requirement has been classified, it can be easily quantified following the theoretical framework that underpins \approach.

Using the same examples as before, ``\texttt{the system should be fast}'' means $\langle\psi_l,\psi_r\rangle=\langle\mathcal{S},\mathcal{S}\rangle$ and $v_{\beta}=N/A$; ``\texttt{The system should response in 2 seconds}'' should be classified as $\langle\psi_l,\psi_r\rangle=\langle\mathcal{E},\mathcal{S}\rangle$ and $v_{\beta}=2$; ``\texttt{The system should response in 5 seconds and ideally less than 2 seconds}'' has $\langle\psi_l,\psi_r\rangle=\langle\mathcal{E},\mathcal{S}\rangle$ and $v_{\beta}=2$ together with $\langle\psi_l,\psi_r\rangle=\langle\mathcal{E},\mathcal{S}\rangle$ and $v_{\beta}=5$.




%% file: figures/example1.tex
\begin{tikzpicture}
    \begin{axis}[
      width=4cm,
    height=4cm,
     axis x line  = bottom,
    axis y line  = left,
    ymax=1.1,
    ymin=0,
    xmax=1.1,
        xlabel=latency (s)/$v$,
        ylabel=$g(v)$,
        xtick={0,1},
          ytick={0,1},
         xticklabels={0,10},
          yticklabels={0,1},
       legend style={at={(0.5,0.95),font=\fontsize{7}{8}\selectfont},
      anchor=north west,legend columns=1}]

 \addplot[no marks,blue,thick,name path=A] plot coordinates {
(0,1)
(1,0)
};

 \addplot[no marks,name path=B] plot coordinates {
(0,0)
(1,0)
};

 \addplot[blue!40,opacity=0.5] fill between[of=A and B];
%
          
       \end{axis}
       \node at (0.5,0.5) {$\psi_1$};
    \end{tikzpicture}

%% file: figures/example2.tex
\begin{tikzpicture}
    \begin{axis}[
      width=4cm,
    height=4cm,
     axis x line  = bottom,
    axis y line  = left,
    ymax=1.1,
    ymin=0,
    xmax=1.1,
     xlabel=latency (s)/$v$,
        ylabel=$g(v)$,
        xtick={0,0.2,1},
          ytick={0,1},
         xticklabels={0,2,10},
          yticklabels={0,1},
       legend style={at={(0.5,0.95),font=\fontsize{7}{8}\selectfont},
      anchor=north west,legend columns=1}]

 \addplot[no marks,blue,thick,name path=A] plot coordinates {
(0,1)
(0.2,1)
};

 \addplot[no marks,blue,thick,name path=B] plot coordinates {
(0.2,1)
(1,0)
};

 \addplot[no marks,name path=C] plot coordinates {
(0,0)
(0.2,0)
};

 \addplot[no marks,name path=D] plot coordinates {
(0.2,0)
(1,0)
};

 \addplot[blue!40,opacity=0.5] fill between[of=A and C];
  \addplot[red!40,opacity=0.5] fill between[of=B and D];
%
          
       \end{axis}
       \node at (0.3,0.5) {$\psi_1$};
       \node at (1,0.5) {$\psi_2$};
    \end{tikzpicture}

%% file: figures/example3.tex
\begin{tikzpicture}
    \begin{axis}[
      width=4cm,
    height=4cm,
     axis x line  = bottom,
    axis y line  = left,
    ymax=1.1,
    ymin=0,
    xmax=1.1,
     xlabel=latency (s)/$v$,
        ylabel=$g(v)$,
        xtick={0,0.2,0.35,0.5,1},
          ytick={0,1},
         xticklabels={0,2,3.5,5,10},
          yticklabels={0,1},
       legend style={at={(0.5,0.95),font=\fontsize{7}{8}\selectfont},
      anchor=north west,legend columns=1}]

 \addplot[no marks,blue,thick,name path=A] plot coordinates {
(0,1)
(0.2,1)
};

 \addplot[no marks,blue,thick,name path=B] plot coordinates {
(0.2,1)
(0.35,0.5)
};

 \addplot[no marks,blue,thick,name path=C] plot coordinates {
(0.35,0.5)
(0.5,0.5)
};

 \addplot[no marks,blue,thick,name path=D] plot coordinates {
(0.5,0.5)
(1,0)
};

 \addplot[no marks,name path=E] plot coordinates {
(0,0)
(0.2,0)
};

 \addplot[no marks,name path=F] plot coordinates {
(0.2,0)
(0.35,0)
};

 \addplot[no marks,name path=G] plot coordinates {
(0.35,0)
(0.5,0)
};

 \addplot[no marks,name path=H] plot coordinates {
(0.5,0)
(1,0)
};

 \addplot[blue!40,opacity=0.5] fill between[of=A and E];
  \addplot[red!40,opacity=0.5] fill between[of=B and F];
  \addplot[teal!40,opacity=0.5] fill between[of=C and G];
  \addplot[orange!40,opacity=0.5] fill between[of=D and H];
%
          
       \end{axis}
         \node at (0.2,0.5) {$\psi_1$};
       \node at (0.6,0.5) {$\psi_2$};
       \node at (1,0.5) {$\psi_3$};
          \node at (1.4,0.5) {$\psi_4$};
    \end{tikzpicture}

%% file: method.tex
\section{Automated Quantification with \approach}
\label{sec:method}

\subsection{Why not ``Learn'' from Data?}


Indeed, our formulated classification appears to fit well with a machine learning classifier. Yet, the insights from Section~\ref{sec:insights} suggest that the simple learning approaches can easily overfit the learned samples, weakening their generalization. The limited available performance requirements data further exacerbate this issue, e.g., a few hundreds compared with tens of thousands or even millions in other software engineering tasks~\cite{DBLP:conf/msr/DabicAB21,DBLP:conf/msr/SpinellisKM20}.


Foundation models like LLM also seems to be universally applicable, but it is ill-fitted for our case because the problem is specifically formed according to strong domain understanding (from \textbf{\textit{Finding 1-2}}), hence the resulted formulation does not align with the general knowledge that LLM is trained for, exacerbating the hallucination issue. Therefore, we opt for a linguistics-induced approach to solve our formulated classification problem.

\subsection{\approach~Workflow}

\begin{figure}[!t]
 \centering
 \includegraphics[width=\columnwidth]{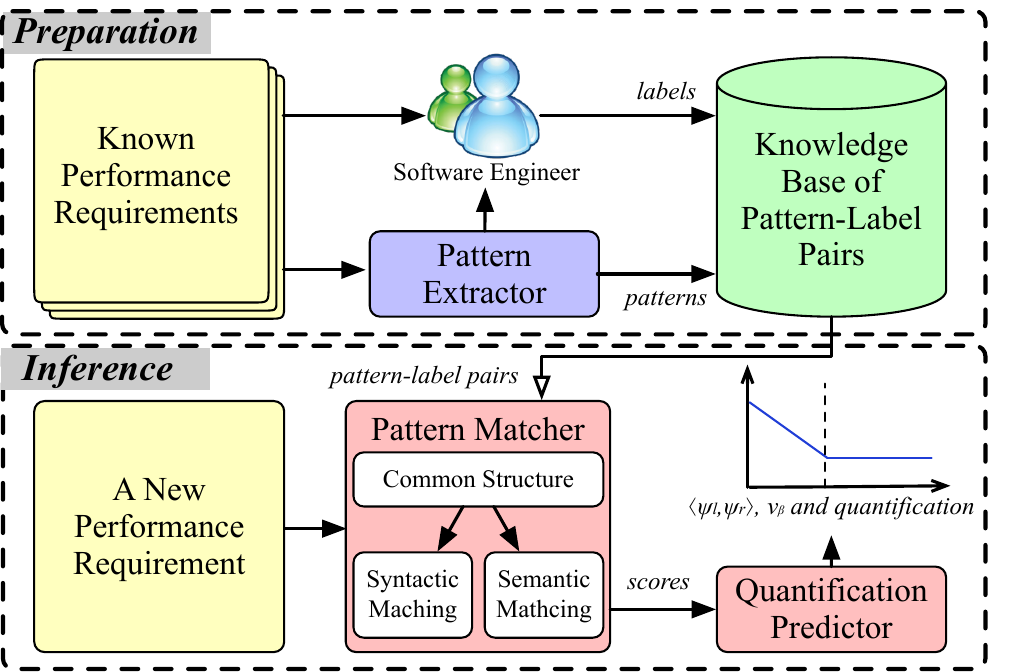}
  \caption{Workflow and architecture of \approach.}
   \label{arch}
  \vspace{-0.3cm}
 \end{figure}

As can be seen from Figure~\ref{arch}, the key idea of \approach~is that we can linguistically infer to what extent a given performance requirement matches with each pattern extracted, and therefore quantifying its preference according to the best match and the theoretical framework. To that end, we design the following components in \approach:

\begin{itemize}
    \item \textbf{Pattern Extractor} automatically extracts the patterns from known performance requirements, which are manually labeled according to the theoretical framework in Section~\ref{sec:theory}. These serve as the knowledge base of linguistic inducement.
    \item \textbf{Pattern Matcher} finds the common structure between a given performance requirement (or a split requirement) and each of the patterns, the match of which is dually scored with respect to syntax and semantic.

    
    \item Finally, \textbf{Quantification Predictor} selects the best-match using the scores, from which infers the label and quantify the requirement via the theoretical framework in Section~\ref{sec:theory}.
\end{itemize}

Only the \textbf{Pattern Extractor} runs in the \textit{preparation phase} while the others are part of the actual \textit{inference} phase.

\subsection{Linguistic Patterns Extraction and Labeling}


Deriving from \textbf{\textit{Findings 1-2}}, we note that there are clear patterns of performance requirements that strongly indicate their implied preferences in classification. As such, in \approach, we form a predefined set of patterns and their corresponding labels, each is represented as ``$pattern \rightarrow label$'. In particular, we found that the complement (e.g., a verb or adverb) and the expectation point (if any) used to describe the concerned subject, together with those describe the expectation point itself, are crucial and highly informative\footnote{Those phrases also strong indicate whether the metric is to be minimized or maximized, e.g., ``\texttt{capable of}'' and ``\texttt{support}'' only appear in requirements for maximizing metrics like throughput. Others like ``\texttt{shall be}'' can be used on any metrics.}. Suppose that we have a requirement ``\texttt{the throughput should support more than 100 requests}'', after manual analysis and labeling, the above means that there are some tolerance for value smaller than the expectation but anything greater than that are equally preferred, i.e., $\langle\mathcal{G},\mathcal{E}\rangle$. Here, the extracted pattern,
together with its label, can be expressed as a pattern-label pair below:
\begin{equation}
\overbrace{
\text{\texttt{more than }} v_{\beta}
}^{\text{\textbf{\textit{pattern}}}}
 \rightarrow 
 \overbrace{
\langle\psi_l,\psi_r\rangle=\langle\mathcal{G},\mathcal{E}\rangle 
}^{\text{\textbf{\textit{label}}}}
\end{equation}
where $v_{\beta}$ is preserved for the expectation point (excluding unit), which might varies depending on different performance requirement. Any requirement that fits with the above patterns can be classified/quantified in the same way. 

\begin{figure}[!t]
 \centering
\includegraphics[width=\columnwidth]{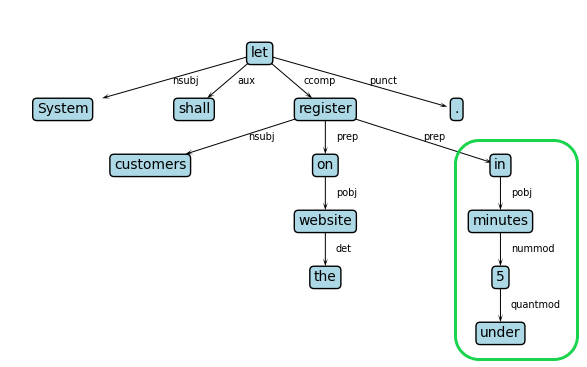}
  \caption{The syntactic dependency tree of a performance requirement. The branch of extracted pattern is highlighted.}
   \label{depTree}
    \vspace{-0.3cm}
 \end{figure}

To automatically extract the patterns, \approach~builds a syntactic dependency tree for the given performance requirement, in which each branch is a grammatical group. The pattern we seek is the branch(es) contains the numeric expectation or phrase that directly describes the concerned condition, subject or action. For example in Figure~\ref{depTree}, for the performance requirement ``\texttt{system shall let customers register on the website in under 5 minutes}'', the highlighted branch, which contains the expectation of ``\texttt{5 minutes}'' that describes the action ``\texttt{register}'' alongside the complement of the expectation ``\texttt{in under}'', is the pattern we wish to extract. 



In \approach, we use linguistic tool \textsc{spaCy}~\cite{spacy} to build the syntactic dependency tree of performance requirements and automatically extract the patterns, which are manually labeled via the quantification/classes from the theoretical framework. Those form a knowledge base of patterns, enabling linguistic-induced quantification.



\subsection{Linguistic Structure Commonality}

Using the patterns, \approach~then parses their linguistically common structure against a given performance requirement\footnote{Requirements with more than one expectations can be easily split using \textsc{spaCy}.}. To that end, \approach~formulates the structure commonality identification as a Longest Common Subsequence (LCS) searching problem---a concept in linguistic analysis~\cite{wang2007all,DBLP:journals/jacm/Maier78}---due to its efficiency and high suitability to our needs. In our case, this means that we find the longest subsequences of identical words/tokens between a given performance requirement and a pattern. In particular, such a common subsequence does not require the words to be consecutive, i.e., the words are common as long as their orders are identical even though they might have differently interleaving words. Note that the $v_{\beta}$ refers to any number regardless of the value, thus there is a LCS in the performance requirement as long as a number follows the last word in the pattern. As such, the actual value of $v_{\beta}$ can be extracted from the match for quantification. For example, 
for an extracted pattern ``\texttt{\textbf{be capable of} supporting $\bm{v_{\beta}}\rightarrow \langle\psi_l,\psi_r\rangle=\langle\mathcal{G},\mathcal{E}\rangle$}'' and given a performance requirement ``\texttt{the product shall \textbf{be capable of} handling the existing \textbf{1000} users}'',  their LCS is ``\texttt{\texttt{be capable of 1000}}'' with a length of $4$, and we know that $v_{\beta}=1000$.

In \approach~we use a dynamic programming solver~\cite{bellman1966dynamic} to find the LCS. The time complexity of this process is $O(m \times n)$, where $m$ and $n$ are the lengths of the given performance requirement and the number of patterns, respectively.

\subsection{Structure-driven Syntactic Matching}


Leveraging on the common structure of LCS between a given performance requirement and the $k$th pattern, \approach~distinguishes their syntactic match via a score $m'_{k,a}$:
\begin{equation}
\label{eq:main}
m'_{k,a} =  {l_{k,lcs} \over l_k}
\end{equation}
whereby $l_{k,lcs}$ and $l_k$ is the length of the LCS and pattern, respectively. A longer pattern is naturally more likely to have longer LCS, thus we prefer a shorter pattern. 


However, since LCS naturally does not consider any information about the closeness between the words in the match, using $m'_{k,a}$ might lead to the identical score for the matches with many different patterns. For example, when scoring the requirement ``\texttt{the product shall {be capable of} handling the existing {1000} users}'' against two patterns ``\texttt{{be capable of} supporting ${v_{\beta}}\rightarrow \langle\psi_l,\psi_r\rangle=\langle\mathcal{G},\mathcal{E}\rangle$}'' and ``\texttt{{shall be} ${v_{\beta}}\rightarrow \langle\psi_l,\psi_r\rangle=\langle\mathcal{G},\mathcal{S}\rangle$}'', the former is more syntactically fitted and contain more similar implied preference. However, the latter will have higher $m'_{k,a}$ score (i.e., $1$) than that of the former (i.e., ${4 \over 5}$). To address this, \approach~computes the syntactic matches via a penalized score $m_{k,a}$:
\begin{equation}
m_{k,a} =  m'_{k,a} \times {l_{k,lcs} \over {\max(l_{k,lcs},l_s)}}, \text{ s.t. } l_s = i_{last} - i_{first} + 1
\end{equation}
where $l_s$ is the distance between the first and last word from the LCS appear in the given performance requirement. In this way, we penalize the matches whose LCSs are more deviated apart in the given performance requirement, since they are less likely to imply similar preferences and hence being less syntactically fitted. With the same example as above, the $l_{k,lcs}$ and $l_s$ with pattern "\texttt{shall be $v_{\beta}$}" are $3$ and $7$, respectively, leading to $m'_{k,a}=1$ and $m_{k,a}={3 \over 7}=0.429$; in contrast, pattern "\texttt{be capable of supporting $v_{\beta}$}" would have $l_{k,lcs}=4$ and $l_s=6$, hence $m'_{k,a}={4 \over 5}$ and $m_{k,a}={4 \over 5} \times {4 \over 6} = 0.533$. Therefore, the latter pattern, which is indeed more syntactically fitted, can be reflected by a higher score for the given performance requirement.

\subsection{Structure-driven Semantic Matching}

While the syntactic analysis is useful, it can hardly handle the semantic information~\cite{DBLP:conf/acl/ZhaoHLB18}. Yet, we cannot directly compare the semantic of requirements with the pattern because the requirement could still contain ``noisy words'' that misleads the latent embedding. For example, in the requirement ``\texttt{Upon the USB being plugged in, the system shall be able to be deployed and operational in {less than 100} minutes}'', the phrase ``\texttt{Upon the USB being plugged in}'' does not contribute to the implied preferences for quantification while ``\texttt{deployed}'' and ``\texttt{operational}'' play similar role in the semantic interpretation, hence unnecessarily influence the pattern machining results: it should be more semantically similar to the pattern ``\texttt{less than $v_{\beta}$}'' than ``\texttt{in under $v_{\beta}$}'', but comparing the entire requirement leads to the opposed result.

Therefore, \approach~adopts the structure information to derive semantic analysis~\cite{DBLP:journals/dke/OlivaSCI11,DBLP:journals/kais/Aygun08}: since the extracted LCS represents the most common part from the requirement analyzed, we compare such a LCS against the pattern for a sematic match via:



\begin{enumerate}
    \item For the $k$th pattern and its extracted LCS, compute their word vectors via \texttt{word2vec}\footnote{We use a highly simple model of only 11MB pre-trained by the \texttt{OntoNotes} corpus.} as $\{\bm{\vec{w}_1},\bm{\vec{w}_2},...,\bm{\vec{w}_i}\}$ and $\{\bm{\vec{w}'_1},\bm{\vec{w}'_2},$ $...,\bm{\vec{w}'_j}\}$, respectively. The corresponding sentences vector $\bm{\vec{v}_{k}}$ and $\bm{\vec{v}_{k,lcs}}$ can be calculated by averaging the word vectors, e.g., $\bm{\vec{v}_{k}} = {{\bm{\vec{w}_1}+\bm{\vec{w}_2}+...+\bm{\vec{w}_i}} \over i}$.
    \item To score the semantic match between the LCS and a pattern, we use the cosine similarity:
    \begin{equation}
m_{k,b} =  \Vert \bm{\vec{v}_{k}} \Vert \Vert \bm{\vec{v}_{k,lcs}} \Vert  \cos \theta 
\end{equation}
\item Repeat (1) until all the patterns are examined.
\end{enumerate}

The pattern with a higher $m_{k,b}$ against the given requirement should be semantically more similar. For example,  when there is a requirement ``\texttt{The system response time for all operations should be under 3 seconds}'', if we only consider the syntactic matching, its $m_{k,a}$ with pattern ``\texttt{all must be}'' and ``\texttt{in under $v_{\beta}$}'' are both $0.667$. However, it is clear that the requirement is more semantically similar to the latter pattern, which can be correctly reflected using the above semantic matching $m_{k,b}$ (via the corresponding LCSs ``\texttt{all be}'' and ``\texttt{under 3}''). \revision{Note that to ensure the overall lightweight nature of \approach, we employ the simplified version of \texttt{word2vec} and utilize the standard averaged word vectors as the representation. This approach proves effective given the concise and succinct nature of the patterns. Indeed, using other methods with higher computational overhead might assure better results.}


The core complexity of the above come from the cosine similarity computation, which depends on the length of requirements. Yet, we found that the semantic matching in \approach~is highly efficient due to the commonly short performance requirements.

\subsection{Classifying/Quantifying via Pattern Label}

The final pattern is selected by finding the highest of dually scored syntactic and semantic match as follows:
\begin{equation}
\argmax w \times m_{k,a} + (1- w) \times  {(m_{k,b}+1) \over 2}
\end{equation}
whereby the scale of $m_{k,b}$ is normalized while $m_{k,a}$ naturally ranges within $[0,1]$. $w$ is the weight that controls the relative importance between syntax and semantic matching, for which we found that $w=0.7$ is an optimal value (see Section~\ref{sec:sen}). Once we identify the best-matched pattern, \approach~then uses the corresponding label and $v_{\beta}$ for classification and quantification. Yet, instead of always directly using the label, there are cases where we need to reverse it, i.e., when the requirement contains negation terms and they are not part of the corresponding LCS. \revision{To detect those cases, we construct a negation lexicon, such as ``\texttt{not}'', ``\texttt{no}'', or ``\texttt{neither}'' \textit{etc}\footnote{A complete list can be found at: \textcolor{blue}{\texttt{\url{https://github.com/ideas-labo/LQPR/blob/main/pattern/negative_word.txt}}}.}, and perform sequential detection of these vocabulary items within the statement. If one of those keywords is found, we reverse the fragment from $\mathcal{S}$ to $\mathcal{G}$ and vice versa in the label. This straightforward strategy proves suitable for concise requirement statements while ensuring high processing efficiency.}

For example, both ``\texttt{the response
time shall be no more than 100 milliseconds}'' and ``\texttt{the throughput shall be more than 200 users}'' have the highest dual score to the pattern ``\texttt{more than} $v_{\beta} \rightarrow \langle\psi_l,\psi_r\rangle=\langle\mathcal{G},\mathcal{E}\rangle$'', but they imply completely opposite preferences, and only the latter requirement can be classified correctly without reversing the label. Since the former has a negative term ``\texttt{no}'', which is not in LCS, and it seeks to minimizing the metric while ``\texttt{no}'' is a strong term that implies nothing worse than the expectation is acceptable, \approach~reverses the corresponding label from $\langle\mathcal{G},\mathcal{E}\rangle$ to $\langle\mathcal{S},\mathcal{E}\rangle$, which matches the needs of response time.

The label (and $v_{\beta}$, if any) can then be directly quantified following the procedure in Section~\ref{sec:theory}. Any previously split requirements with their labels can also be combined again and jointly quantified.

%% file: study.tex
\section{Experimental Design}
\label{sec:study}

\subsection{Research Questions}

In this work, we examine several research questions (RQs):

\begin{itemize}

\item \textbf{RQ1:} How does \approach~perform against state-of-the-art on performance requirements within a dataset?
\item \textbf{RQ2:} To what extent can \approach~generalize to performance requirements across datasets?
\item \textbf{RQ3:} What are the contributions of each design in \approach?
\item \textbf{RQ4:} What is the efficiency of \approach?
\end{itemize}

All the experiments are conducted on a high-performance server with Ubuntu 20.04.1 LTS, Intel(R) Xeon(R) Platinum 8480$+$ with 224 CPU cores and 500GB memory.

\subsection{Dataset}
\label{subsec:dataset}



\revision{The datasets are selected with the following criteria:}
\begin{itemize}
    \item \revision{The dataset must include formal elicitation processes, i.e., there are no informal requirements posted on online platforms such as \texttt{StackOverflow} (the student projects in \textsc{Promise} also underwent some levels of requirement elicitation).}  
    \item \revision{It must contain a sufficient number of performance-related and quantifiable requirements ($\geq 10$ after splitting), which are extracted using the procedure elaborated in Section~\ref{sec:emprical}.}  
\end{itemize}
\revision{The results are shown in Table~\ref{tb:dataset}.} Note that a performance requirement with more than one expectation points is split into multiple requirements. \revision{To further enrich our experiments, we use a LLM (\texttt{GPT-4} in this case) to generate a synthetic dataset for testing (i.e., \textsc{LLM-Gen}). We do so by prompting the LLM with 20 examples randomly chosen from \textsc{Promise} (excluding those used for training/patterning), based on each of which we ask it to generate five new but diverse requirements (with one expectation point).}
{Finally, all performance requirements are manually labeled by the authors, who are experienced software engineers, according to the theoretical framework in Section~\ref{sec:theory}.}

\input{tables/dataset}

\subsection{Learning-based Approaches}

We compare \approach~with several state-of-the-art learning-based approaches commonly used for requirement analytics.

\subsubsection{Statistical Machine Learning Classifiers with Texts Vectorization}

We compare Naive Bayes (\texttt{NB}) and $k$ Nearest Neighbor ($k$\texttt{NN} with $k=5$ as the default), each paired with Term Frequency-Inverse Term Frequency (\texttt{TF-IDF}) and Bag-of-Words (\texttt{BoW}) for texts vectorization as commonly used in the literature~\cite{DBLP:journals/entropy/CanedoM20,DBLP:journals/tse/ZhaoXW24,DBLP:conf/re/DalpiazDAC19}.

\subsubsection{Encoder-only LLMs}

We examine \texttt{BERT}, a widely used pre-trained encoder-only LLMs for classifying requirements~\cite{DBLP:journals/corr/abs-1907-11692,DBLP:conf/naacl/DevlinCLT19,DBLP:journals/infsof/AlhoshanFZ23}. Here, we compare two variants: the classic fine-tuned \texttt{BERT}~\cite{DBLP:journals/corr/abs-1907-11692,DBLP:conf/naacl/DevlinCLT19} and \texttt{BERT} with zero-shot learning (\texttt{ZSL}~\cite{DBLP:journals/infsof/AlhoshanFZ23}). The former is pre-trained with Wikipedia data and fine-tuned using samples from the \textsc{Promise} dataset. The latter is pre-trained with requirement data without fine-tuning as proposed by Alhoshan et al.~\cite{DBLP:journals/infsof/AlhoshanFZ23}. \finalrevision{We examine \texttt{PRCBERT}~\cite{DBLP:conf/kbse/LuoXXS22}, a \texttt{RoBERT}-based model pre-trained on the \textsc{Promise} dataset, designed for standard requirement-type classification tasks. We use the same pre-trained model and protocol from the authors for fine-tuning.}

\subsubsection{Decoder-only LLMs}

We study top-3 performed decoder-only LLMs from \texttt{TogetherAI}~\cite{aicloud}---a well-known LLM leaderboard---namely \texttt{Gemma-27B}~\cite{gemma}, \texttt{Deepseek-67B}~\cite{deepseek}, and \texttt{Llama-8B}~\cite{llama}. As part of the prompt for each prediction, we perform in-context learning by providing the decoder-only LLMs with 10 examples of correctly labeled/quantified performance requirements (covering all classes) according to our theoretical framework\footnote{An exampled prompt can be found at: \textcolor{blue}{\texttt{\url{https://github.com/ideas-labo/LQPR/blob/main/prompt/question.txt}}}.}, and ask them to infer the label of the given example in the same format.

\subsection{Metrics and Statistical Test}


In the evaluation, we use widely adopted metrics, i.e., precision, recall, and F1-score~\cite{dinga2019beyond}. In particular, since we are dealing with an imbalanced data \revision{multi-class classification} problem, we use the weighted version of the metrics~\cite{DBLP:conf/ease/Fu0L021,DBLP:conf/icse/PanBXLL23} ($wP$, $wR$, and $wF_1$), i.e., the value of each label is linearly averaged and weighted according to the label's proportion in the dataset. Notably, since the common interpretation for the requirement engineering is that only a metric value greater than 0.8 can lead to useful classification~\cite{DBLP:journals/infsof/AlhoshanFZ23,DBLP:conf/re/Williams18}, we do not see \approach~as achieving a sufficiently good result when it is worse than 0.8 even if it is significantly superior to the others.

The experiments are repeated 30 runs and to ensure validity, we use the Scott-Knott test~\cite{mittas2012ranking}---a clustering algorithm based on the statistical differences of approaches---to assess statistical significance. Assuming three approaches $A$, $B$, and $C$, the Scott-Knott test may yield two groups: $\{A, B\}$ with rank 1 and $\{C\}$ with rank 2, meaning that $A$ and $B$ are statistically similar but they are both significantly better than $C$. \revision{Note that Scott-Knott test has internally used effect size to cluster/rank the approaches; those with small effect sizes would have been clustered into the same rank.}

%% file: tables/dataset.tex
\begin{table}[t!]
\centering

\caption{Datasets and projects studied.}

\label{tb:dataset}
\adjustbox{max width=\columnwidth}{
\begin{tabular}{llll}
 \toprule

\textbf{Dataset} & \textbf{$\#$ Projects} & \textbf{$\#$ Perf. Requirements} & \textbf{Source}  \\

\midrule

\textsc{Promise}~\cite{promise}&15&259&real-world\\
\textsc{PURE}~\cite{DBLP:conf/re/FerrariSG17}&79&23&real-world\\
Shaukat et al.~\cite{DBLP:conf/cse/ShaukatNZ18}&4&15&real-world\\
\revision{\textsc{Functional-Quality}~\cite{DBLP:journals/corr/abs-2504-16768}}&\revision{5}&\revision{10}&\revision{real-world}\\
\textsc{LLM-Gen}&N/A&100&synthetic\\

\bottomrule
\end{tabular}
}
    \vspace{-0.3cm}
\end{table}

%% file: results.tex
\section{Evaluation}
\label{sec:exp}

\subsection{RQ1: Inferring for Within Dataset}

\subsubsection{Method}
To answer \textbf{RQ1}, we seek to evaluate the ability of \approach~in inferring and quantifying requirements from the same dataset sampled for pattern extraction. To that end, \revision{we perform bootstrapping without replacement by randomly sampling around $2\over3$ data, i.e., 170 out of 259 performance requirements, from the \textsc{Promise} dataset for 30 runs (with different seeds). These are the samples for training the statistical machine learning classifiers and \texttt{BERT}; and for \approach~to extract patterns\footnote{We found that the patterns do not change much across the runs; an exampled list can be found at: \textcolor{blue}{\texttt{\url{https://github.com/ideas-labo/LQPR/blob/main/pattern/patterns.txt}}}.}. The remaining samples in \textsc{Promise} are used for testing.}

\subsubsection{Results}

From Table~\ref{tb:rq1}, we note that decoder-only LLMs perform similarly to those statistical machine learning approaches, both of which are worse than encoder-only LLMs like \texttt{BERT}; this is possible, since statistical machine learning might easily overfit while the decoder-only LLMs are often more suitable for generative tasks; the \texttt{BERT}, on the other hand, is specifically designed for classification, which is what we need. The \texttt{ZSL} leads to even worse results than classic machine learning approaches, because it is trained using general requirement data which might contain samples that are non-performance related, hence causing noises. In contrast, \approach~performs remarkably well, outperforming all learning-based approaches on nearly all cases. The results are all above 0.8 and \approach~is generally ranked as the sole best. This suggests that the specifically designed linguistic inducement in \approach~is better-suited to the formulated classification and is practically useful. We say:

\keystate{
\textit{\textbf{RQ1:} \approach~performs considerably better than the state-of-the-art, including LLMs, for performance requirements collected in the same dataset as those used for patterns extraction.}
}

\input{tables/rq1}

\subsection{RQ2: Inferring for Cross Datasets}

\input{tables/rq2}

\subsubsection{Method}

\textbf{RQ2} seeks to examine \approach~on inferring performance requirements from completely \revision{unseen} datasets for generalizbility. \revision{To that end, we use the same bootstrapping as \textbf{RQ1} to select 170 samples from \textsc{Promise} for pattern extraction and training; the approaches are then tested on all samples from the other datasets.}

\subsubsection{Results}

Table~\ref{tb:rq2} shows that \approach~performs considerably better in general, being ranked the sole best for 75\% (9 out of 12) cases. Compared with \textbf{RQ1}, the relative differences do not change much, but we see that most approaches tend to perform slightly better, as there are more cases with results greater than 0.8. This is due to the natural differences between the datasets: \textsc{Promise} contains more diverse performance requirements hence it is often more representative; each of the others, although collected/generated by distinct protocols, might involve many samples from certain categories that have been well-captured by \textsc{Promise}. Overall, we conclude that:

Table~\ref{tb:rq2} shows that \approach~performs considerably better in general, being ranked the sole best for 75\% (9 out of 12) cases. Compared with \textbf{RQ1}, the relative differences do not change much, but we see that most approaches tend to perform slightly better, as there are more cases with results greater than 0.8. This is due to the natural differences between the datasets: \textsc{Promise} contains more diverse performance requirements hence it is often more representative; each of the others, although collected/generated by distinct protocols, might involve many samples from certain categories that have been well-captured by \textsc{Promise}. \finalrevision{We also see that the decoder-LLM approaches have no deviation, this is because although there are variations in the generated text across the runs, the extracted classification labels are consistent.} Overall, we conclude that:

\keystate{
\textit{\textbf{RQ2:} \approach~better generalizes to \revision{unseen} datasets than state-of-the-art for all cases, on 88\% of which it is ranked the sole best.}
}

\subsection{RQ3: Ablation Study}
\label{sec:rq3}

\input{tables/rq3}

\subsubsection{Method}

For \textbf{RQ3}, we remove each of the key designs in turn to verify its contribution to \approach. This has led to three variants:

\begin{itemize}
    \item \approach\texttt{-L}: we ignore negative terms without label reversal.
    \item \approach\texttt{-se}: \approach~with semantic matching only.
    \item \approach\texttt{-sy}: \approach~with syntactic matching only.
\end{itemize}

\revision{We use the same training/extraction and testing samples as in \textbf{RQ1} and \textbf{RQ2} for the corresponding datasets.}
\subsubsection{Results}

From Table~\ref{tb:rq3}, we see that \approach~performs the best overall. \approach\texttt{-L} contributes to the results significantly more than the others on the datasets that involve many negative terms \revision{(i.e., \textsc{Promise} and \textsc{LLM-GEN})}, which, if not handled correctly, would certainly cause wrong inference. It is clear that ignoring any of the syntax and semantic matching could lead to harmful implication (e.g., for the Shaukat et al. and \textsc{PURE} datasets), and hence combining both in the linguistic induced analysis is important. 

\revision{We also notice that for \textsc{Promise} and \textsc{LLM-Gen} where the requirements are of similar structure, using only the syntactic information can maximize its benefit (without being negatively impacted by the semantic part), hence \approach\texttt{-sy} performs similar to \approach. However, for the \textsc{PURE} and Shaukat et al. datasets, where the syntactic structure is less common, the full \approach~performs significantly better. The above proves the robustness of \approach.}

In summary, we say:

\keystate{
\textit{\textbf{RQ3:} All the key designs in \approach~are indeed beneficial.}
}

\subsection{RQ4: Efficiency}


\subsubsection{Method}

To verify the efficiency in \textbf{RQ4}, we study the clock time and memory resource required for training/fine-tuning and inference by all approaches. We omit the pattern extraction/labeling since this is a common process with manual reasoning. \revision{Again, the training/extraction and testing are the same as the previous RQs.}

\subsubsection{Results}

Figure~\ref{fig:rq4} shows that, unsurprisingly, most LLMs, albeit do not need downstream training, consume a much higher memory and power while incur longer runtime upon inference. \revision{\texttt{BERT} and \texttt{PRCBERT} are the LLMs that needs downstream fine-tuning, and hence them also consumes a significant amount of resources and clock time.} The statistical machine learning approaches are often highly efficient for inferences but still require a considerable amount of time for training merely 170 samples, which can be devastating when the model needs to be updated/used for, e.g., runtime self-adaptation~\cite{DBLP:conf/se/WeynsGAABBBSGLL24}. \approach, in contrast, is much more lightweight: it does not require any downstream training while incurring little inference overhead with up to two orders of efficiency improvement than the others on both space and time. That is to say:

\begin{figure}[t!]
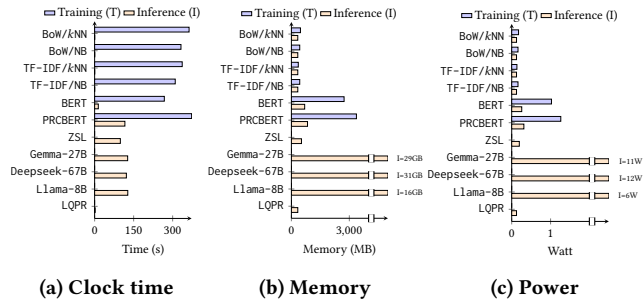

\centering
\subfloat[Clock time]{\includestandalone[width=0.3145\columnwidth]{figures/new-rq4-1}}
~\hspace{-0.15cm}
\subfloat[Memory]{\includestandalone[width=0.35\columnwidth]{figures/new-rq4-2}}
~\hspace{-0.15cm}
\subfloat[Power]{\includestandalone[width=0.35\columnwidth]{figures/new-rq4-3}}

\caption{Efficiency on clock time, memory, and power consumption. The training refers to the downstream task training/fine-tuning of 170 samples, excluding any pre-training. Inference time and its consumption are the average for one sample only. The power is estimated from the hardware resources consumed based on a prior work~\cite{bouza2023estimate}.}  
\label{fig:rq4}
  \vspace{-0.3cm}
\end{figure}

\keystate{
\textit{\textbf{RQ4:} The superior performance of \approach~comes with little cost---at least two orders more efficient than the others such as LLMs.}
}

%% file: tables/rq1.tex
\begin{table}[t!]
\centering

\setlength{\tabcolsep}{3mm}
\caption{Classifying and quantifying performance requirements from the same known dataset. We report on the mean and standard deviation (SD) over 30 runs. $r$ denotes Scott-Knott rank; \setlength{\fboxsep}{1.5pt}\colorbox{orange!30}{orange cells} indicate the best on a metric. The mean results better than 0.8 are highlighted in \textbf{bold}.}

\label{tb:rq1}
\adjustbox{max width=\columnwidth}{
\begin{tabular}{l|ll|ll|ll}
 \toprule

\multirow{2}{*}{\textbf{Approach}} & \multicolumn{2}{|c}{\textbf{$\bm{wP}$}} & \multicolumn{2}{|c}{\textbf{$\bm{wR}$}} & \multicolumn{2}{|c}{\textbf{$\bm{wF_1}$}} \\

\cmidrule{2-7}

&$\bm{r}$&\textbf{Mean (SD)}&$\bm{r}$&\textbf{Mean (SD)}&$\bm{r}$&\textbf{Mean (SD)}\\

\midrule

\texttt{BoW/NB} & 3 & 0.680 (0.063) & 2 & 0.730 (0.048) & 3 & 0.680 (0.058) \\
\texttt{BoW/$k$NN} & 3 & 0.690 (0.050) & 3 & 0.680 (0.047) & 3 & 0.650 (0.051) \\
\texttt{TF-IDF/NB} & 4 & 0.610 (0.046) & 2 & 0.720 (0.042) & 3 & 0.650 (0.047) \\
\texttt{TF-IDF/$k$NN} & 3 & 0.680 (0.042) & 3 & 0.690 (0.037) & 3 & 0.670 (0.041) \\
\texttt{BERT} & \textbf{2} & \textbf{0.840 (0.051)} &\cellcolor{orange!30}\textbf{1} & \cellcolor{orange!30}\textbf{0.850 (0.036)} & \textbf{2} & \textbf{0.830 (0.043)} \\
\texttt{\revision{PRCBERT}} & \revision{4} & \revision{0.603 (0.051)} & \revision{2} & \revision{0.727 (0.042)} & \revision{3} & \revision{0.648 (0.091)} \\
\texttt{ZSL} & 4 & 0.590 (0.158) & 5 & 0.390 (0.044) & 5 & 0.400 (0.047) \\
\texttt{Gemma-27B} & 4 & 0.600 (3.128) & 4 & 0.590 (3.059) & 4 & 0.590 (2.666) \\
\texttt{Deepseek-67B} & 4 & 0.590 (2.660) & 4 & 0.590 (2.983) & 4 & 0.590 (3.115) \\
\texttt{Llama-8B} & 4 & 0.600 (2.518) & 4 & 0.590 (2.202) & 4 & 0.600 (2.988) \\
\texttt{LQPR} & \cellcolor{orange!30}\textbf{1} & \cellcolor{orange!30}\textbf{0.861 (0.024)} & \cellcolor{orange!30}\textbf{1} & \cellcolor{orange!30}\textbf{0.858 (0.026)} & \cellcolor{orange!30}\textbf{1} & \cellcolor{orange!30}\textbf{0.853 (0.025)} \\

\bottomrule
\end{tabular}
}
  \vspace{-0.3cm}
\end{table}

%% file: tables/rq2.tex
\begin{table}[t!]
\centering

\setlength{\tabcolsep}{3mm}
\caption{Classifying and quantifying performance requirements from unforeseen datasets. The format is as Table~\ref{tb:rq1}.}

\label{tb:rq2}
\adjustbox{max width=\columnwidth}{
\begin{tabular}{l|ll|ll|ll}
    \toprule

\multirow{2}{*}{\textbf{Approach}} & \multicolumn{2}{|c}{\textbf{$\bm{wP}$}} & \multicolumn{2}{|c}{\textbf{$\bm{wR}$}} & \multicolumn{2}{|c}{\textbf{$\bm{wF_1}$}} \\
\cmidrule{2-7}
&$\bm{r}$&\textbf{Mean (SD)}&$\bm{r}$&\textbf{Mean (SD)}&$\bm{r}$&\textbf{Mean (SD)}\\
\midrule
\multicolumn{7}{c}{\cellcolor{gray!30}\textbf{\textit{\textsc{PURE} Dataset}}}\\
\midrule

\texttt{BoW/NB} & \textbf{2} & \textbf{0.830 (0.056)} & \textbf{3} & \textbf{0.810 (0.043)} & \textbf{2} & \textbf{0.820 (0.044)} \\
\texttt{BoW/$k$NN} & 4 & 0.750 (0.068) & 5 & 0.690 (0.073) & 5 & 0.700 (0.066) \\
\texttt{TF-IDF/NB} & 3 & 0.790 (0.035) & \textbf{3} & \textbf{0.820 (0.033)} & \textbf{3} & \textbf{0.800 (0.032)} \\
\texttt{TF-IDF/$k$NN} & \textbf{2} & \textbf{0.820 (0.042)} & 4 & 0.780 (0.047) & 3 & 0.790 (0.041) \\
\texttt{BERT} & \textbf{2} & \textbf{0.820 (0.033)} & \textbf{2} & \textbf{0.850 (0.039)} & \textbf{2} & \textbf{0.830 (0.042)} \\
\revision{\texttt{PRCBERT}} & \revision{5} & \revision{0.678 (0.053)} & \revision{5} & \revision{0.743 (0.082)} & \revision{5} & \revision{0.714 (0.039)}\\
\texttt{ZSL} & 5 & 0.249 (0.000) & 7 & 0.197 (0.000) & 6 & 0.220 (0.000) \\
\texttt{Gemma-27B} & \textbf{3} & \textbf{0.802 (0.000)} & \textbf{3} & \textbf{0.808 (0.000)} & 4 & 0.775 (0.000) \\
\texttt{Deepseek-67B} & 4 & 0.785 (0.000) & 5 & 0.705 (0.000) & 4 & 0.760 (0.000) \\
\texttt{Llama-8B} & 3 & 0.792 (0.000) & 6 & 0.634 (0.000) & 5 & 0.649 (0.000) \\
\texttt{LQPR} & \cellcolor{orange!30}\textbf{1} & \cellcolor{orange!30}\textbf{0.951 (0.000)} & \cellcolor{orange!30}\textbf{1} & \cellcolor{orange!30}\textbf{0.947 (0.000)} & \cellcolor{orange!30}\textbf{1} & \cellcolor{orange!30}\textbf{0.946 (0.000)} \\

\midrule

\multicolumn{7}{c}{\cellcolor{gray!30}\textbf{\textit{Shaukat et al. Dataset}}}\\
\midrule

\texttt{BoW/NB} & \textbf{3} & \textbf{0.800 (0.051)} & \textbf{2} & \textbf{0.820 (0.046)} & \textbf{2} & \textbf{0.800 (0.045)} \\
\texttt{BoW/$k$NN} & 4 & 0.770 (0.042) & 5 & 0.630 (0.103) & 5 & 0.640 (0.097) \\
\texttt{TF-IDF/NB} & 3 & 0.790 (0.043) & \textbf{2} & \textbf{0.820 (0.034)} & 3 & 0.790 (0.034) \\
\texttt{TF-IDF/$k$NN} & 4 & 0.750 (0.059) & 4 & 0.740 (0.056) & 4 & 0.740 (0.051) \\
\texttt{BERT} & 3 & 0.790 (0.041) & \textbf{2} & \textbf{0.830 (0.034)} & \textbf{2} & \textbf{0.810 (0.036)} \\
\revision{\texttt{PRCBERT}} & \revision{4} & \revision{0.732 (0.061)} & \revision{4} & \revision{0.728 (0.040)} & \revision{4} & \revision{0.736 (0.059)} \\
\texttt{ZSL} & 5 & 0.210 (0.000) & 6 & 0.168 (0.000) & 6 & 0.187 (0.000) \\
\texttt{Gemma-27B} & \textbf{2} & \textbf{0.846 (0.000)} & \textbf{3} & \textbf{0.807 (0.000)} & \textbf{2} & \textbf{0.823 (0.000)} \\
\texttt{Deepseek-67B} & \textbf{2} & \textbf{0.857 (0.000)} & 4 & 0.709 (0.000) & 3 & 0.761 (0.000) \\
\texttt{Llama-8B} & \textbf{3} & \textbf{0.806 (0.000)} & 5 & 0.672 (0.000) & 5 & 0.674 (0.000) \\
\texttt{LQPR} & \cellcolor{orange!30}\textbf{1} & \cellcolor{orange!30}\textbf{1.000 (0.000)} & \cellcolor{orange!30}\textbf{1} & \cellcolor{orange!30}\textbf{1.000 (0.000)} & \cellcolor{orange!30}\textbf{1} & \cellcolor{orange!30}\textbf{1.000 (0.000)} \\

\midrule

\multicolumn{7}{c}{\cellcolor{gray!30}\textbf{\textit{\revision{\textsc{Functional-Quality} Dataset}}}}\\
\midrule

\texttt{BoW/NB} & 6 & 0.256 (0.276) & 6 & 0.350 (0.086) & 7 & 0.220 (0.135) \\
\texttt{BoW/$k$NN} & 5 & 0.559 (0.166) & 4 & 0.580 (0.127) & 6 & 0.513 (0.110) \\
\texttt{TF-IDF/NB} & \textbf{2} & \textbf{0.918 (0.051)} & \cellcolor{orange!30}\textbf{1} & \cellcolor{orange!30}\textbf{0.877 (0.086)} & \textbf{2} & \textbf{0.880 (0.083)} \\
\texttt{TF-IDF/$k$NN} & \textbf{2} & \textbf{0.871 (0.120)} & 2 & \textbf{0.803 (0.130)} & 3 & \textbf{0.816 (0.128)} \\
\texttt{BERT} & 4 & 0.681 (0.227) & 2 & 0.793 (0.123) & 4 & 0.717 (0.178) \\
\revision{\texttt{PRCBERT}} & \revision{3} & \revision{0.772 (0.455)} & \revision{2} & \revision{\textbf{0.825 (0.350)}} & \revision{3} & \revision{0.784 (0.431)} \\
\texttt{ZSL} & \cellcolor{orange!30}\textbf{1} & \cellcolor{orange!30}\textbf{1.000 (0.000)} & 5 & 0.500 (0.000) & 5 & 0.611 (0.000) \\
\texttt{Gemma-27B} & 3 & \textbf{0.802 (0.000)} & 2 & \textbf{0.808 (0.000)} & 3 & 0.775 (0.000) \\
\texttt{Deepseek-67B} & 3 & 0.785 (0.000) & 3 & 0.705 (0.000) & 3 & 0.760 (0.000) \\
\texttt{Llama-8B} & 3 & 0.792 (0.000) & 4 & 0.634 (0.000) & 5 & 0.649 (0.000) \\
\texttt{LQPR} & \cellcolor{orange!30}\textbf{1} & \cellcolor{orange!30}\textbf{1.000 (0.000)} & \cellcolor{orange!30}\textbf{1} & \cellcolor{orange!30}\textbf{0.900 (0.000)} & \cellcolor{orange!30}\textbf{1} & \cellcolor{orange!30}\textbf{0.946 (0.000)} \\

\midrule

\multicolumn{7}{c}{\cellcolor{gray!30}\textbf{\textit{\textsc{LLM-Gen} Dataset}}}\\
\midrule

\texttt{BoW/NB} & 4 & 0.780 (0.035) & 4 & 0.680 (0.054) & 4 & 0.720 (0.049) \\
\texttt{BoW/$k$NN} & 4 & 0.740 (0.048) & 5 & 0.380 (0.051) & 5 & 0.360 (0.073) \\
\texttt{TF-IDF/NB} & 4 & 0.780 (0.026) & 3 & 0.750 (0.075) & 3 & 0.750 (0.068) \\
\texttt{TF-IDF/$k$NN} & 4 & 0.760 (0.046) & 4 & 0.690 (0.088) & 4 & 0.710 (0.076) \\
\texttt{BERT} & \textbf{2} & \textbf{0.870 (0.022)} & \cellcolor{orange!30}\textbf{1} & \cellcolor{orange!30}\textbf{0.880 (0.016)} & \textbf{2} & \textbf{0.870 (0.018)} \\
\revision{\texttt{PRCBERT}} & \revision{3} & \revision{0.841 (0.091)} & \revision{2} & \revision{0.857 (0.035)} & \revision{2} & \revision{0.828 (0.077)} \\
\texttt{ZSL} & 5 & 0.215 (0.000) & 6 & 0.190 (0.000) & 6 & 0.202 (0.000) \\
\texttt{Gemma-27B} & \textbf{2} & \textbf{0.901 (0.000)} & \textbf{2} & \textbf{0.840 (0.000)} & \textbf{2} & \textbf{0.864 (0.000)} \\
\texttt{Deepseek-67B} & \textbf{3} & \textbf{0.863 (0.000)} & 3 & 0.710 (0.000) & 3 & {0.770} (0.000) \\
\texttt{Llama-8B} & \textbf{2} & \textbf{0.879 (0.000)} & 4 & 0.700 (0.000) & 4 & 0.685 (0.000) \\
\texttt{LQPR} & \cellcolor{orange!30}\textbf{1} & \cellcolor{orange!30}\textbf{0.945 (0.000)} & \cellcolor{orange!30}\textbf{1} & \cellcolor{orange!30}\textbf{0.880 (0.000)} & \cellcolor{orange!30}\textbf{1} & \cellcolor{orange!30}\textbf{0.906 (0.000)} \\
\bottomrule
\end{tabular}
}
  \vspace{-0.3cm}
\end{table}

%% file: tables/rq3.tex
\begin{table}[t!]
\centering

\setlength{\tabcolsep}{4mm}
\caption{Ablation analysis of \approach~by excluding different designs one at a time. Other formats are the same as Table~\ref{tb:rq1}.}

\label{tb:rq3}
\adjustbox{max width=\columnwidth}{
\begin{tabular}{l|ll|ll|ll}
\toprule

\multirow{2}{*}{\textbf{Approach}} & \multicolumn{2}{|c}{\textbf{$\bm{wP}$}} & \multicolumn{2}{|c}{\textbf{$\bm{wR}$}} & \multicolumn{2}{|c}{\textbf{$\bm{wF_1}$}} \\
\cmidrule{2-7}
&$\bm{r}$&\textbf{Mean (SD)}&$\bm{r}$&\textbf{Mean (SD)}&$\bm{r}$&\textbf{Mean (SD)}\\
\midrule
    
\multicolumn{7}{c}{\cellcolor{gray!30}\textbf{\textit{\textsc{Promise} Dataset}}}\\
\midrule
\approach\texttt{-L} & 3 & 0.787 (0.032) & 2 & 0.752 (0.040) & 2 & 0.747 (0.042) \\
\approach\texttt{-se} & \textbf{2} & \textbf{0.827 (0.033)} & 3 & 0.460 (0.033) & 3 & 0.563 (0.033) \\
\approach\texttt{-sy} & \cellcolor{orange!30}\textbf{1} & \cellcolor{orange!30}\textbf{0.875 (0.021)} & \cellcolor{orange!30}\textbf{1} & \cellcolor{orange!30}\textbf{0.862 (0.024)} & \cellcolor{orange!30}\textbf{1} & \cellcolor{orange!30}\textbf{0.865 (0.023)} \\
\approach & \cellcolor{orange!30}\textbf{1} & \cellcolor{orange!30}\textbf{0.861 (0.024)} & \cellcolor{orange!30}\textbf{1} & \cellcolor{orange!30}\textbf{0.858 (0.026)} & \cellcolor{orange!30}\textbf{1} & \cellcolor{orange!30}\textbf{0.853 (0.025)} \\

\midrule

\multicolumn{7}{c}{\cellcolor{gray!30}\textbf{\textit{\textsc{PURE} Dataset}}}\\
\midrule

\approach\texttt{-L} & \textbf{2} & \textbf{0.895 (0.000)} & \textbf{2} & \textbf{0.895 (0.000)} & \textbf{2} & \textbf{0.895 (0.000)} \\
\approach\texttt{-se} & \cellcolor{orange!30}\textbf{1} & \cellcolor{orange!30}\textbf{0.952 (0.000)} & 3 & 0.473 (0.000) & 3 & 0.588 (0.000) \\
\approach\texttt{-sy} & \textbf{3} & \textbf{0.857 (0.000)} & \textbf{2} & \textbf{0.895 (0.000)} & \textbf{2} & \textbf{0.870 (0.000)} \\
\approach & \cellcolor{orange!30}\textbf{1} & \cellcolor{orange!30}\textbf{0.951 (0.000)} & \cellcolor{orange!30}\textbf{1} & \cellcolor{orange!30}\textbf{0.947 (0.000)} & \cellcolor{orange!30}\textbf{1} & \cellcolor{orange!30}\textbf{0.946 (0.000)} \\

\midrule

\multicolumn{7}{c}{\cellcolor{gray!30}\textbf{\textit{Shaukat et al. Dataset}}}\\
\midrule

 \approach\texttt{-L} & \textbf{2} & \textbf{0.950 (0.000)} & \textbf{2} & \textbf{0.933 (0.000)} & \textbf{2} & \textbf{0.937 (0.000)} \\
\approach\texttt{-se} & \textbf{2} & \textbf{0.942 (0.000)} & 3 & 0.600 (0.000) & 4 & 0.694 (0.000) \\
\approach\texttt{-sy} & \textbf{3} & \textbf{0.872 (0.000)} & \textbf{2} & \textbf{0.933 (0.000)} & \textbf{3} & \textbf{0.901 (0.000)} \\
\approach & \cellcolor{orange!30}\textbf{1} & \cellcolor{orange!30}\textbf{1.000 (0.000)} & \cellcolor{orange!30}\textbf{1} & \cellcolor{orange!30}\textbf{1.000 (0.000)} & \cellcolor{orange!30}\textbf{1} & \cellcolor{orange!30}\textbf{1.000 (0.000)} \\

\midrule

\multicolumn{7}{c}{\cellcolor{gray!30}\textbf{\textit{\revision{\textsc{Functional-Quality} Dataset}}}}\\
\midrule

\texttt{LQPR-L} & \textbf{2} & \textbf{0.925 (0.000)} & \textbf{2} & \textbf{0.800 (0.000)} & \textbf{2} & \textbf{0.840 (0.000)} \\
\texttt{LQPR-se} & 3 & 0.700 (0.000) & 3 & 0.300 (0.000) & 3 & 0.420 (0.000) \\
\texttt{LQPR-sy} & \textbf{2} & \textbf{0.925 (0.000)} & \cellcolor{orange!30}\textbf{1} & \cellcolor{orange!30}\textbf{0.900 (0.000)} & \cellcolor{orange!30}\textbf{1} & \cellcolor{orange!30}\textbf{0.903 (0.000)} \\
\texttt{LQPR} & \cellcolor{orange!30}\textbf{1} & \cellcolor{orange!30}\textbf{1.000 (0.000)} & \cellcolor{orange!30}\textbf{1} & \cellcolor{orange!30}\textbf{0.900 (0.000)} & \cellcolor{orange!30}\textbf{1} & \cellcolor{orange!30}\textbf{0.946 (0.000)} \\

\midrule
    
\multicolumn{7}{c}{\cellcolor{gray!30}\textbf{\textit{\textsc{LLM-Gen} Dataset}}}\\
\midrule

\approach\texttt{-L} & 2 & 0.761 (0.000) & 3 & 0.580 (0.000) & 3 & 0.617 (0.000) \\
\approach\texttt{-se} & \cellcolor{orange!30}\textbf{1} & \cellcolor{orange!30}\textbf{0.952 (0.000)} & \textbf{2} & \textbf{0.750 (0.000)} & \textbf{2} & \textbf{0.827 (0.000)} \\
\approach\texttt{-sy} &  \cellcolor{orange!30}\textbf{1} &  \cellcolor{orange!30}\textbf{0.942 (0.000)} &  \cellcolor{orange!30}\textbf{1} &  \cellcolor{orange!30}\textbf{0.870 (0.000)} &  \cellcolor{orange!30}\textbf{1} &  \cellcolor{orange!30}\textbf{0.896 (0.000)} \\
\approach & \cellcolor{orange!30}\textbf{1} & \cellcolor{orange!30}\textbf{0.945 (0.000)} & \cellcolor{orange!30}\textbf{1} & \cellcolor{orange!30}\textbf{0.880 (0.000)} & \cellcolor{orange!30}\textbf{1} & \cellcolor{orange!30}\textbf{0.906 (0.000)} \\
        
\bottomrule
\end{tabular}
}
  \vspace{-0.2cm}
\end{table}

%% file: figures/new-rq4-1.tex


    \begin{tikzpicture}
        \begin{axis}[
            xbar,
            axis x line  = bottom,
            axis y line  = left  ,
             reverse legend,
            bar width=.12cm,
            width=4cm,
            height=6.5cm,
              xtick={0,150,300},
               enlarge y limits=0.07,
            xlabel={Time (s)},
            symbolic y coords={\texttt{LQPR},\texttt{Llama-8B},\texttt{Deepseek-67B},\texttt{Gemma-27B},\texttt{ZSL},\texttt{PRCBERT},\texttt{BERT},\texttt{TF-IDF/NB},\texttt{TF-IDF/$k$NN},\texttt{BoW/NB},\texttt{BoW/$k$NN}},
            ytick=data,
            xmin=0,
            legend style={at={(0.3,1.1)},draw=none,fill=none,anchor=north,legend columns=2},
            legend cell align={left},
            legend image code/.code={
        \draw [#1] (0cm,-0.05cm) rectangle (0.25cm,0.05cm); }
        ]
      

   \addplot+[fill=orange!20,draw=black
            ] plot coordinates {
              (1.696,\texttt{LQPR}) +- (0.256,0)
            (127.4865,\texttt{Llama-8B}) +- (1.55847504,0) 
            (122.0732,\texttt{Deepseek-67B}) +- (2.504538792,0)
            (127.0165,\texttt{Gemma-27B}) +- (0.002891220616,0)
            (98.8954,\texttt{ZSL}) +- (2.115061001,0)
             (116.36581,\texttt{PRCBERT}) +- (0.195667,0)
            (14.16712606,\texttt{BERT}) +- (0.1267098741,0)
            (0.02502834797,\texttt{TF-IDF/NB}) +- (0.003357434094,0)
            (0.04073073864,\texttt{TF-IDF/$k$NN}) +- (0.004916353175,0) 
            (0.2313415289,\texttt{BoW/NB}) +- (0.003328942283,0) 
            (0.01532664299,\texttt{BoW/$k$NN}) +- (0.003557387056,0) 
        };

   \addplot+[fill=blue!20,draw=black
            ] plot coordinates {
            (0,\texttt{LQPR}) +- (0,0)
            (0,\texttt{Llama-8B}) +- (0,0) 
            (0,\texttt{Deepseek-67B}) +- (0,0)
            (0,\texttt{Gemma-27B}) +- (0,0)
            (0,\texttt{ZSL}) +- (0,0)
             (373,\texttt{PRCBERT}) +- (28.3,0)
            (268,\texttt{BERT}) +- (15.8,0)
            (310,\texttt{TF-IDF/NB}) +- (3.84,0)
            (336.8,\texttt{TF-IDF/$k$NN}) +- (6.20,0) 
             (332,\texttt{BoW/NB}) +- (13.8,0) 
              (363.2,\texttt{BoW/$k$NN}) +- (4.75,0) 
        };

         \legend{Inference (I),Training (T)}

        \end{axis}










    \end{tikzpicture}

%% file: figures/new-rq4-2.tex


    \begin{tikzpicture}
        \begin{axis}[
            xbar,
            axis x line  = bottom,
            axis y line  = left  ,
             reverse legend,
            bar width=.12cm,
            width=4cm,
            height=6.5cm,
               xtick={0,3000},
               enlarge y limits=0.07,
            xlabel={Memory (MB)},
            symbolic y coords={\texttt{LQPR},\texttt{Llama-8B},\texttt{Deepseek-67B},\texttt{Gemma-27B},\texttt{ZSL},\texttt{PRCBERT},\texttt{BERT},\texttt{TF-IDF/NB},\texttt{TF-IDF/$k$NN},\texttt{BoW/NB},\texttt{BoW/$k$NN}},
            ytick=data,
            xmin=0,
          legend style={at={(0.3,1.1)},draw=none,fill=none,anchor=north,legend columns=2},
            legend cell align={left},
            legend image code/.code={
        \draw [#1] (0cm,-0.05cm) rectangle (0.25cm,0.05cm); }
        ]
      

   \addplot+[fill=orange!20,draw=black
            ] plot coordinates {
             (331.6,\texttt{LQPR}) +- (5.680375574,0)
            (5000,\texttt{Llama-8B}) +- (0,0) 
            (5000,\texttt{Deepseek-67B}) +- (0,0)
            (5000,\texttt{Gemma-27B}) +- (0,0)
            (521.6,\texttt{ZSL}) +- (4.948624949,0)
             (834.7,\texttt{PRCBERT}) +- (5.763581,0)
            (688.04,\texttt{BERT}) +- (7.325328965,0)
            (331.9,\texttt{TF-IDF/NB}) +- (5.300943312,0)
            (331.6,\texttt{TF-IDF/$k$NN}) +- (6.257440016,0) 
            (331.7,\texttt{BoW/NB}) +- (5.716448004,0) 
            (330.6,\texttt{BoW/$k$NN}) +- (6.569288817,0)  
        };

   \addplot+[fill=blue!20,draw=black
            ] plot coordinates {
           (0,\texttt{LQPR}) +- (0,0)
            (0,\texttt{Llama-8B}) +- (0,0) 
            (0,\texttt{Deepseek-67B}) +- (0,0)
            (0,\texttt{Gemma-27B}) +- (0,0)
            (0,\texttt{ZSL}) +- (0,0)
            (3367.5,\texttt{PRCBERT}) +- (9.456,0)
             (2728.3,\texttt{BERT}) +- (5.96,0)
            (432,\texttt{TF-IDF/NB}) +- (3.88,0)
            (357,\texttt{TF-IDF/$k$NN}) +- (5.35,0) 
             (432,\texttt{BoW/NB}) +- (3.20,0) 
              (469,\texttt{BoW/$k$NN}) +- (6.64,0) 
        };

         \legend{Inference (I),Training (T)}

        \end{axis}
      \draw[thick,double,double distance=3pt](2,-0.1)--(2,0.1);
\draw[thick,double,double distance=3pt](2,0.55)--(2,0.73);
\draw[thick,double,double distance=3pt](2,0.98)--(2,1.18);
\draw[thick,double,double distance=3pt](2,1.4)--(2,1.6);


 \node at (2.97,0.65) {\scriptsize I=16GB};

 \node at (2.97,1.08) {\scriptsize I=31GB};

 \node at (2.97,1.5) {\scriptsize I=29GB};






    \end{tikzpicture}

%% file: figures/new-rq4-3.tex


    \begin{tikzpicture}
        \begin{axis}[
            xbar,
            axis x line  = bottom,
            axis y line  = left  ,
             reverse legend,
            bar width=.12cm,
            width=4cm,
            height=6.5cm,
               xtick={0,1},
               enlarge y limits=0.07,
            xlabel={Watt},
            symbolic y coords={\texttt{LQPR},\texttt{Llama-8B},\texttt{Deepseek-67B},\texttt{Gemma-27B},\texttt{ZSL},\texttt{PRCBERT},\texttt{BERT},\texttt{TF-IDF/NB},\texttt{TF-IDF/$k$NN},\texttt{BoW/NB},\texttt{BoW/$k$NN}},
            ytick=data,
            xmin=0,
          legend style={at={(0.3,1.1)},draw=none,fill=none,anchor=north,legend columns=2},
            legend cell align={left},
            legend image code/.code={
        \draw [#1] (0cm,-0.05cm) rectangle (0.25cm,0.05cm); }
        ]
      
   \addplot+[fill=orange!20,draw=black
            ] plot coordinates {
             (0.1238,\texttt{LQPR}) +- (5.680375574,0)
            (2.5,\texttt{Llama-8B}) +- (0,0) 
            (2.5,\texttt{Deepseek-67B}) +- (0,0)
            (2.5,\texttt{Gemma-27B}) +- (0,0)
            (0.1958,\texttt{ZSL}) +- (0.004948624949,0)
             (0.3130125,\texttt{PRCBERT}) +- (0.004948624949,0)
            (0.2580,\texttt{BERT}) +- (0.007325328965,0)
            (0.1238,\texttt{TF-IDF/NB}) +- (0.005300943312,0)
            (0.1238,\texttt{TF-IDF/$k$NN}) +- (0.006257440016,0) 
            (0.1238,\texttt{BoW/NB}) +- (0.005716448004,0) 
            (0.1238,\texttt{BoW/$k$NN}) +- (0.006569288817,0)  
        };

   \addplot+[fill=blue!20,draw=black
            ] plot coordinates {
           (0,\texttt{LQPR}) +- (0,0)
            (0,\texttt{Llama-8B}) +- (0,0) 
            (0,\texttt{Deepseek-67B}) +- (0,0)
            (0,\texttt{Gemma-27B}) +- (0,0)
            (0,\texttt{ZSL}) +- (0,0)
              (1.2628125,\texttt{PRCBERT}) +- (0.004948624949,0)
             (1.0231,\texttt{BERT}) +- (0.00596,0)
            (0.1620,\texttt{TF-IDF/NB}) +- (0.00388,0)
            (0.1339,\texttt{TF-IDF/$k$NN}) +- (0.00535,0) 
             (0.1620,\texttt{BoW/NB}) +- (0.00320,0) 
              (0.1758,\texttt{BoW/$k$NN}) +- (0.00664,0) 
        };

         \legend{Inference (I),Training (T)}

        \end{axis}
        \draw[thick,double,double distance=3pt](2,-0.1)--(2,0.1);
\draw[thick,double,double distance=3pt](2,0.55)--(2,0.73);
\draw[thick,double,double distance=3pt](2,0.98)--(2,1.18);
\draw[thick,double,double distance=3pt](2,1.4)--(2,1.6);


 \node at (2.9,0.65) {\scriptsize I=6W};

 \node at (2.97,1.08) {\scriptsize I=12W};

 \node at (2.97,1.5) {\scriptsize I=11W};






    \end{tikzpicture}

%% file: discussion.tex
\section{Discussion}
\label{sec:discussion}

\subsection{Why \approach~Surpasses Statistical Learning?}

The most common misclassified examples for statistical machine learning approaches like \texttt{TF-IDF/$k$NN} are the following:
\sta{
\centering
``{The system shall have a downtime of {at most 10} minutes per year.}''
}
\sta{
\centering
``{The device shall consume {at most 50} watts of power in operation.}''
}
Clearly, while they refer to different performance metrics, the correct labels of both should be $\langle\mathcal{S},\mathcal{E}\rangle$, indicating that better than the expectation point is preferred to some extents and does not accept anything worse off. However, \texttt{TF-IDF/$k$NN} has classified the above as $\langle\mathcal{E},\mathcal{S}\rangle$ and $\langle\mathcal{G},\mathcal{E}\rangle$, respectively. This is because the machine learning approaches tend to over-fit all the (non-important) vocabularies from the samples trained, hence harming the generalization. \approach, in contrast, classifies both correctly thanks to dually scored structure-driven matching with ``\texttt{{at most} ${v_{\beta}}\rightarrow \langle\psi_l,\psi_r\rangle=\langle\mathcal{S},\mathcal{E}\rangle$}''.

\subsection{Why \approach~Outperforms LLMs?}
A common mistake that LLMs made is on those requirements without expectation. For example:
\sta{
\centering
``{The software shall generate reports in an acceptable time.}''
}
The correct class should be $\langle\mathcal{S},\mathcal{S}\rangle$ as without expectation, the general knowledge is that for time-related performance, the smaller, the better. Yet, LLMs incorrectly infer it as $\langle\mathcal{G},\mathcal{G}\rangle$, which have a completely opposed meaning such that longer time is preferred due to the confusion caused by hallucination in their reasoning. \approach~correctly classifies that using the ``\texttt{in an acceptable time}'' pattern. The other examples are requirements such as: 
\sta{
\centering
``{The server shall synchronize with the backup system every 2 hours.}''
}
The correct quantification is $\langle\mathcal{G},\mathcal{S}\rangle$, since neither lower nor higher than 2 hours are preferred. Yet, LLMs have mistakenly classified that as $\langle\mathcal{E},\mathcal{S}\rangle$, meaning less than 2 hours are equally preferred and higher than 2 hours can be tolerated. This is due to the LLMs cannot fully understand the formulated classification problem for requirement quantification, hence they hallucinate the preferences based on the general knowledge that shorter time is better. \approach~can better capture the above via the pattern ``\texttt{every $v_{\beta}$}''.

The above is because the classification problem we seek to address is formulated according to strong domain knowledge, therefore LLMs cannot gain benefits from the general understanding that they were pre-trained, even with proper in-context learning. Beside, the fact that performance requirements are often short restricts LLMs to obtain sufficient information and signals. \approach~prevents the above issue by extracting the strong linguistic information from the requirements and matching it with prior understandings.

\begin{figure}[t!]
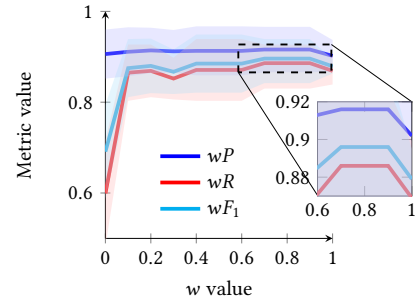

\centering
\includestandalone[width=0.65\columnwidth]{figures/double-weight}
\caption{Overall sensitivity of \approach~to $w$ on all datasets.}  
\label{fig:sen}
  \vspace{-0.3cm}
\end{figure}

\subsection{Sensitivity to $w$}
\label{sec:sen}


$w$ controls the contributions between syntactic and semantic matching. \revision{Figure ~\ref{fig:sen} shows the overall results by averaging those for all datasets with testing samples as in \textbf{RQ1-2}}. Clearly, the $w$ closer to either $0$ or $1$ reduces the performance, hence both parts are important. $w \in [0.7,0.9]$ can lead to optimality, meaning that the syntactic information should be preferred more than its semantic counterpart. This is because the syntax is often more important for matching short texts such as performance requirements.

\subsection{Limitations of \approach}

A shortcoming of \approach~is that it relies on the patterns to ensure accuracy, therefore, although rare, there exist requirements that have been completely missed by any known patterns. For example:
\sta{
\centering
``{\revision{The system design should ensure stability even when serving 500 concurrent active users.}}''
}
\revision{This performance requirement does not contain information close to any known patterns of \approach, especially because the way that the preference is expressed as \texttt{``when serving 500 concurrent active users''}. As a result, \approach~has failed to classify it correctly.}  However, the patterns can be easily enriched with more examples. 

\revision{The other limitation is that \approach~cannot directly predict the $v_{\beta}$ if such a value is implicit. However, it serves as a foundation towards a solution. For example, one might provide several possible values of the implicit $v_{\beta}$(in different statements) for \approach~to make predictions, and then examine the inferred quantification while changing some requirements for ``what-if'' analysis. Through multiple iterations/interactions, \approach~can help refine the $v_{\beta}$ value and eventually reach one's real expectation.}


%% file: figures/double-weight.tex
\begin{tikzpicture}
    \begin{axis}[
      axis x line  = bottom,
            axis y line  = left  ,
        width=5cm, height=5cm, 
        xlabel={$w$ value},
        ylabel={Metric value},
        xmin=0, xmax=1,
        ymin=0.5, ymax=1, 
        xtick={0, 0.2, 0.4, 0.6, 0.8, 1.0}, 
        grid=minor,
        legend style={draw=none,at={(0.65,0.05)}, anchor=south east, legend cell align=left}
    ]

    \addplot[
        blue,
        mark=none,
        ultra thick
    ] coordinates {
        (0.0, 0.906) (0.1, 0.911) (0.2, 0.914) (0.3, 0.912) (0.4, 0.913) (0.5, 0.913) (0.6, 0.913) (0.7, 0.916) (0.8, 0.916) (0.9, 0.916) (1.0, 0.902)
    };
    \addlegendentry{$wP$}

    \addplot[
        red,
        mark=none,
        ultra thick
    ] coordinates {
        (0.0, 0.599) (0.1, 0.865) (0.2, 0.869) (0.3, 0.852) (0.4, 0.871) (0.5, 0.871) (0.6, 0.871) (0.7, 0.886) (0.8, 0.886) (0.9, 0.886) (1.0, 0.869)
    };
    \addlegendentry{$wR$}

    \addplot[
        cyan,
        mark=none,
        ultra thick
    ] coordinates {
        (0.0, 0.691) (0.1, 0.875) (0.2, 0.88) (0.3, 0.867) (0.4, 0.885) (0.5, 0.885) (0.6, 0.885) (0.7, 0.896) (0.8, 0.896) (0.9, 0.896) (1.0, 0.879)
    };
    \addlegendentry{$wF_1$}

    \addplot[
        fill=blue!30, 
        draw=none,
        fill opacity=0.3 
    ] coordinates {
        (0.0, 0.852) (0.1, 0.857) (0.2, 0.864) (0.3, 0.865) (0.4, 0.859) (0.5, 0.859) (0.6, 0.859) (0.7, 0.866) (0.8, 0.866) (0.9, 0.866) (1.0, 0.867)
        (1.0, 0.937) (0.9, 0.966) (0.8, 0.966) (0.7, 0.966) (0.6, 0.967) (0.5, 0.967) (0.4, 0.967) (0.3, 0.959) (0.2, 0.964) (0.1, 0.965) (0.0, 0.96)
    } \closedcycle;

    \addplot[
        fill=red!30, 
        draw=none,
        fill opacity=0.3 
    ] coordinates {
        (0.0, 0.482) (0.1, 0.803) (0.2, 0.811) (0.3, 0.806) (0.4, 0.803) (0.5, 0.803) (0.6, 0.803) (0.7, 0.83) (0.8, 0.83) (0.9, 0.83) (1.0, 0.839)
        (1.0, 0.899) (0.9, 0.927) (0.8, 0.927) (0.7, 0.927) (0.6, 0.939) (0.5, 0.939) (0.4, 0.939) (0.3, 0.898) (0.2, 0.927) (0.1, 0.927) (0.0, 0.716)
    } \closedcycle;

    \addplot[
        fill=cyan!30, 
        draw=none,
        fill opacity=0.3 
    ] coordinates {
        (0.0, 0.587) (0.1, 0.811) (0.2, 0.82) (0.3, 0.818) (0.4, 0.821) (0.5, 0.821) (0.6, 0.821) (0.7, 0.842) (0.8, 0.842) (0.9, 0.842) (1.0, 0.863)
        (1.0, 0.895) (0.9, 0.939) (0.8, 0.939) (0.7, 0.939) (0.6, 0.949) (0.5, 0.949) (0.4, 0.949) (0.3, 0.916) (0.2, 0.94) (0.1, 0.939) (0.0, 0.795)
    } \closedcycle;

    \end{axis}

\draw[black, thick, dashed] (2,2.5) rectangle (3.4,2.92);
\draw (2,2.5) -- (3.2,0.635);
\draw (3.4,2.92) -- (4.62,2.05);

      \begin{axis}[
        width=3cm, height=3cm, 
        xmin=0.6, xmax=1,
        ymin=0.87, ymax=0.92,
        xtick={0.6, 0.8, 1.0}, 
        grid=minor,
        at={(90,22.5)},
        legend style={draw=none,at={(0.95,0.05)}, anchor=south east, legend cell align=left}
    ]

    \addplot[
        blue,
        mark=none,
        ultra thick
    ] coordinates {
        (0.0, 0.906) (0.1, 0.911) (0.2, 0.914) (0.3, 0.912) (0.4, 0.913) (0.5, 0.913) (0.6, 0.913) (0.7, 0.916) (0.8, 0.916) (0.9, 0.916) (1.0, 0.902)
    };

    \addplot[
        red,
        mark=none,
        ultra thick
    ] coordinates {
        (0.0, 0.599) (0.1, 0.865) (0.2, 0.869) (0.3, 0.852) (0.4, 0.871) (0.5, 0.871) (0.6, 0.871) (0.7, 0.886) (0.8, 0.886) (0.9, 0.886) (1.0, 0.869)
    };

    \addplot[
        cyan,
        mark=none,
        ultra thick
    ] coordinates {
        (0.0, 0.691) (0.1, 0.875) (0.2, 0.88) (0.3, 0.867) (0.4, 0.885) (0.5, 0.885) (0.6, 0.885) (0.7, 0.896) (0.8, 0.896) (0.9, 0.896) (1.0, 0.879)
    };

    \addplot[
        fill=blue!30, 
        draw=none,
        fill opacity=0.3 
    ] coordinates {
        (0.0, 0.852) (0.1, 0.857) (0.2, 0.864) (0.3, 0.865) (0.4, 0.859) (0.5, 0.859) (0.6, 0.859) (0.7, 0.866) (0.8, 0.866) (0.9, 0.866) (1.0, 0.867)
        (1.0, 0.937) (0.9, 0.966) (0.8, 0.966) (0.7, 0.966) (0.6, 0.967) (0.5, 0.967) (0.4, 0.967) (0.3, 0.959) (0.2, 0.964) (0.1, 0.965) (0.0, 0.96)
    } \closedcycle;

    \addplot[
        fill=red!30, 
        draw=none,
        fill opacity=0.3 
    ] coordinates {
        (0.0, 0.482) (0.1, 0.803) (0.2, 0.811) (0.3, 0.806) (0.4, 0.803) (0.5, 0.803) (0.6, 0.803) (0.7, 0.83) (0.8, 0.83) (0.9, 0.83) (1.0, 0.839)
        (1.0, 0.899) (0.9, 0.927) (0.8, 0.927) (0.7, 0.927) (0.6, 0.939) (0.5, 0.939) (0.4, 0.939) (0.3, 0.898) (0.2, 0.927) (0.1, 0.927) (0.0, 0.716)
    } \closedcycle;

    \addplot[
        fill=cyan!30, 
        draw=none,
        fill opacity=0.3 
    ] coordinates {
        (0.0, 0.587) (0.1, 0.811) (0.2, 0.82) (0.3, 0.818) (0.4, 0.821) (0.5, 0.821) (0.6, 0.821) (0.7, 0.842) (0.8, 0.842) (0.9, 0.842) (1.0, 0.863)
        (1.0, 0.895) (0.9, 0.939) (0.8, 0.939) (0.7, 0.939) (0.6, 0.949) (0.5, 0.949) (0.4, 0.949) (0.3, 0.916) (0.2, 0.94) (0.1, 0.939) (0.0, 0.795)
    } \closedcycle;

    \end{axis}
\end{tikzpicture}

%% file: threat.tex
\section{Threats to Validity}
\label{sec:threats}


\textbf{Internal validity:} \revision{The only parameter of \approach~is $w$ for which we have empirically set the optimal value based on observation; this might need to be investigated on a case-by-case basis.} For other approaches, we use the default or common parameter settings. The size of patterns/training samples is determined pragmatically, which might not be optimal, but is reasonable for our datasets.

\textbf{Construct validity:} We apply widely used metrics: recall, precision, and F1-score, which are weighted to deal with our imbalanced datasets, with Soctt-Knott test; the quality of classification directly determines the accuracy of quantification. Yet, unintended programming errors or misconsiderations are always possible. 

\textbf{External validity:} The threats to the external validity can raise from the datasets studied. We have considered the most complete and readily available datasets with eligible performance requirements in the field, together with a LLM-generated synthetic one, covering diverse real-world projects. \revision{Yet, we agree that, due to the limited availability of requirements data that fits our needs, those datasets might not be the strongest representatives. The compared approaches might not be the optimal ones, since we are targeting a newly formulated problem, and it is difficult to find directly comparable approaches, but adapting general ones to our contexts.}

%% file: related.tex
\section{Related Work}
\label{sec:related}

\textbf{Requirements formalization:} Approaches for formalizing requirements exist, such as \texttt{RELAX}~\cite{DBLP:journals/re/WhittleSBCB10} and \texttt{FLAGS}~\cite{DBLP:conf/re/BaresiPS10}, which provide formal notations to specify vague requirements. Eckhardt et al.~\cite{DBLP:conf/re/EckhardtVFM16} also present an approach for summarizing the patterns in performance requirements. However, unlike \approach, they have not provided a holistic framework for quantification of performance requirements and rely on manual analysis to formalize the requirements, which could be time-consuming and error-prone. \texttt{AutoRELAX}~\cite{DBLP:journals/ese/FredericksDC14} is an automated extension of \texttt{RELAX}. Yet, their goal is to automatically change a manually pre-defined quantification of requirement to resolve conflicts at runtime, which requires expensive measurements of running system. \approach, in contrast, automatically quantifies performance requirements from natural texts at design time.


\textbf{Requirements analytics with statistical machine learning:} Requirement statements can be classified and analyzed by statistical machine learning~\cite{DBLP:journals/entropy/CanedoM20,DBLP:journals/tse/ZhaoXW24,DBLP:conf/re/DalpiazDAC19}. Among others, Canedo et al.~\cite{DBLP:journals/entropy/CanedoM20} leverage several learners, such as \texttt{Support Vector Machine}, to pair with either \texttt{TF-IDF} or \texttt{BoW} to classify requirements related to different quality aspects of the software systems. Dalpiaz et al.~\cite{DBLP:conf/re/DalpiazDAC19} use syntactic analysis and lexical analysis methods to reduce the dimensionality of the text embedding, which then paired with a conditional judgment algorithm similar to a \texttt{Decision Tree} to classify requirement types, which is also targeted by \revision{Shakeri et al. ~\cite{DBLP:conf/re/AbadKGGRS17} using POS tagging, entity normalization, and temporal expression standardization, together with several machine learning algorithms.} In contrast, \approach~is guided by linguistics knowledge, both syntactically and semantically, to automatically quantify performance requirements in a new theoretical framework.

\finalrevision{\textbf{Natural Language Processing (NLP) for requirements analytics:} For requirements defect detection, Tjong and Berry's SREE tool identifies requirement ambiguities through syntactic analysis~\cite{DBLP:conf/refsq/TjongB13}, while Ferrari et al.~\cite{DBLP:journals/ese/FerrariGRTBFG18} use NLP pattern matching for defect detection in the railway domain; others extract requirement terms based on lexical analysis~\cite{DBLP:journals/jss/AguileraB90};  mine requirements from application reviews~\cite{DBLP:conf/re/MaalejN15}; or transform textual requirements into UML models~\cite{DBLP:journals/tosem/YueBL15}. Yet, none of those can fit the task of requirement quantification.}

\revision{\textbf{LLM for requirements analytics:} The nature of requirements makes them fit well with LLMs. Alhoshan et al.~\cite{DBLP:journals/infsof/AlhoshanFZ23} leverage the \texttt{BERT} pre-trained with requirements data using the zero-shot learning paradigm to classify requirements. Similarly, Hey et al. ~\cite{DBLP:conf/re/HeyKKT20} propose \texttt{NoRBERT}, a method that fine-tunes \texttt{BERT} for the classification of requirement types. Luo et al. ~\cite{DBLP:conf/kbse/LuoXXS22} present the \texttt{PRCBERT} method that is based on \texttt{RoBERTa}. \texttt{PRCBERT} outperforms models like \texttt{NoRBERT} on datasets such as \textsc{Promise} and demonstrates excellent zero-shot performance by integrating a self-learning strategy. Li et al. ~\cite{DBLP:journals/access/LiZLW22} proposed the \texttt{DBGAT} model, integrating \texttt{BERT} and graph attention networks to capture syntactic structure features of requirements through dependency parse trees. \finalrevision{For the decoder-only LLMs, Manal and Reem~\cite{DBLP:journals/corr/abs-2509-13868} tested the effectiveness of prompt-based LLMs (such as GPT) for requirements classification on datasets like \textsc{Promise}.}}

Our work differs from the above in that we formulate a new problem of performance requirement quantification, aiming to automatically quantify the satisfaction function given an elicited requirement statement, which has not been addressed before. Drawing on the observations from performance requirements, we design \approach~as a highly specialized, simpler alternative over the complex ones, tailored to those observations and problem formulated. Common NLP/LLM-based requirement analyses more or less directly leverages on the readily powerful models without or with some amendments. As such, we follow a different technical route.

Further, \approach~do not use informal requests mined from platforms like \texttt{StackOverflow} (e.g., in \texttt{PRCBERT}), as they typically contain user-generated content and the validity cannot be guaranteed. In contrast, \approach~focuses on formally elicited requirements, which inherently contain stronger domain knowledge. Compared to tasks dealing with informal text, these differences can result in text data with unique structures/patterns/concepts and pose specific challenges for automated analysis in requirement quantification.

%% file: conclusion.tex
\section{Conclusion}
\label{sec:con}

This papers proposes a new theoretical framework that formulate the quantification of performance requirements as a classification problem, deriving form empirical insights. We embed the framework within \approach, an automated approach that classifies/quantifies performance requirements based on linguistics knowledge and dual scoring based on observed characteristics. We show that, compared with state-of-the-art approaches such as LLMs, \approach~achieves remarkably better results (being ranked as the sole best for 11 out of 15 cases) with two orders less overhead in general.

\approach~can benefit various performance-related downstream tasks, e.g., configuration tuning~\cite{DBLP:conf/icse/XiongChen25,DBLP:journals/corr/abs-2112-07303,DBLP:conf/sigsoft/0001L24,DBLP:conf/sigsoft/0001L21}, performance prediction~\cite{gong2024dividable,DBLP:conf/sigsoft/Gong023,DBLP:journals/pacmse/Gong024}, and self-adapting systems~\cite{DBLP:journals/tosem/ChenLBY18,DBLP:conf/wcre/Chen22,DBLP:conf/seams/Chen22}. More importantly, our work demonstrates a case of \textit{``light over heavy''}: for software engineering problems that exhibit strong patterns and characteristics, such as performance requirements quantification, specialized and light approach can be preferred over the general, but heavy LLM-driven approaches. This urges the community to take a step back when automating software engineering tasks in the LLM era.

